\documentclass[letterpaper, abstracton, DIV11, 12pt, titlepage=false]{scrartcl}

\usepackage[ruled]{algorithm2e}

\SetAlFnt{\small}
\SetAlCapFnt{\small}
\SetAlCapNameFnt{\small}
\SetAlCapHSkip{0pt}
\IncMargin{-\parindent}
\let\chapter\undefined

\usepackage[usenames,dvipsnames]{xcolor}
\usepackage{footnote}
\usepackage{enumerate}
\usepackage[round]{natbib}
\usepackage[english]{babel}
\usepackage{amsfonts}
\usepackage{amsmath}
\usepackage{amssymb}
\usepackage{booktabs}
\usepackage{cancel}
\usepackage{amsthm}
\usepackage{multirow}
\usepackage{mathabx}
\usepackage{dsfont}
\usepackage{graphicx}
\usepackage{tikz}
\usepackage{pifont}
\usepackage{bm}
\usepackage{wrapfig}
\usepackage{microtype}
\usepackage{setspace}
\usepackage[colorlinks=true,bookmarks=true, pdftex, citecolor = black, linkcolor = black,urlcolor = black]{hyperref}

\onehalfspace

\DisableLigatures[f]{encoding = *, family = * }

\def\bf{\normalfont\bfseries}

\changenotsign

\definecolor{darkgreen}{RGB}{40,150,70}

\newcommand{\URBI}{\text{URBI}}
\newcommand{\URBIr}{\text{URBI($r$)}}
\newcommand{\BM}{\text{BM}}
\newcommand{\ABM}{\text{ABM}}
\newcommand{\f}{f}
\newcommand{\g}{g}

\newtheorem{claim}{Claim}

\theoremstyle{plain}
\newtheorem{theorem}{Theorem}
\newtheorem{lemma}{Lemma}
\newtheorem{proposition}{Proposition}
\newtheorem{corollary}{Corollary}

\theoremstyle{definition}
\newtheorem{definition}{Definition}

\newtheorem{example}{Example}
\newtheorem{axiom}{Axiom}
\newcommand{\ourrep}{}

\theoremstyle{remark}
\newtheorem{remark}{Remark}

\newcommand{\nosep}{\setlength{\itemsep}{0pt}}

\begin{document}

{\setstretch{1}
\title{%
\LARGE{%
Partial Strategyproofness: Relaxing Strategy- proofness for the Random Assignment Problem%
}%
\thanks{%
First version: January 13, 2014;
University of Zurich, Switzerland.
Email: \{mennle, seuken\}@ifi.uzh.ch.
For updates see www.ifi.uzh.ch/ce/publications/PSP.pdf.
We would like to thank (in alphabetical order)
Daniel Ab\"{a}cherli, 
Atila Abdulkadiro\u{g}lu, 
Haris Aziz, 
Ivan Balbuzanov, 
Craig Boutilier, 
Eric Budish, 
Gabriel Carroll, 
Lars Ehlers, 
Katharina Huesmann, 
Bettina Klaus, 
Flip Klijn, 
Fuhito Kojima, 
Eric Maskin, 
Antonio Miralles, 
Bernardo Moreno, 
Thayer Morrill, 
David Parkes, 
Baharak Rastegari, 
Alvin Roth, 
Steffen Schuldenzucker, 
Arunava Sen, 
Tayfun S\"{o}nmez, 
William Thomson, 
and Utku \"{U}nver, 
as well as various anonymous reviewers 
for their thoughtful comments and suggestions. 
We are also thankful for the feedback we received from participants at the following events:
EC '13,
UECE Lisbon Meetings '13,
CIREQ Matching Conference '14,
EC '14,
SCW '14,
IWGTS '14,
ESWC '15,
GAMES '16.
Any errors remain our own.
Some of the ideas presented in this paper were also described in a two-page abstract that was published in the conference proceedings of EC'14; see \citep{MennleSeuken2014_ECExtAbs_AnAxiomaticApproachToCharacterizingAndRelaxingStrategyproofnessOfOneSidedMatchingMechanisms}.
This research was supported by the Hasler Foundation under grant \#12078 and the SNSF (Swiss National Science Foundation) under grant \#156836.}}
\author{%
Timo Mennle \\ University of Zurich
\and Sven Seuken \\ University of Zurich }
\date{July 17, 2020}
\maketitle

\begin{abstract}
We present \emph{partial strategyproofness}, a new, relaxed notion of strategyproofness for studying the incentive properties of non-strategyproof assignment mechanisms.
Informally, a mechanism is partially strategyproof if it makes truthful reporting a dominant strategy for those agents whose preference intensities differ sufficiently between any two objects.
We demonstrate
that partial strategyproofness is axiomatically motivated and yields a parametric 
measure for ``how strategyproof'' an assignment mechanism is. 
We apply this new concept to derive novel insights about the incentive properties of the probabilistic serial mechanism and different variants of the Boston mechanism.
\end{abstract}

}

\newpage
\section{Introduction}
\label{SEC:INTRODUCTION}
The assignment problem is concerned with the allocation of indivisible objects to self-interested
agents who have private preferences over these objects.
Monetary transfers are not permitted, which makes this problem different from auctions and other problems that involve payments. 
In practice, assignment problems often arise in situations that are of great importance to people's lives; for example, when assigning students to seats at 
public schools \citep{AbdulkadirogluSoenmez2003SchoolChoiceAMechanismDesignApproach},
graduates to entry level positions \citep{Roth1984TheEvolutionOfTheLaborMarketForMedicalInternsAndResidentsACaseStudyInGameTheory},
or tenants to subsidized housing \citep{AbdulkadirogluSoenmez1998RandomSerialDictatorshipAndTheCoreFromRandomEndowmentsInHouseAllocationProblems}.

In this paper, we study \emph{ordinal assignment mechanisms}, which are mechanisms for the assignment problem that take preference orders over objects as input.
As mechanism designers, we care specifically about incentives for truthtelling under the mechanisms we design.
A mechanism is \emph{strategyproof} if truthtelling is a dominant strategy equilibrium.
Participating in a strategyproof mechanism is simple for the agents because it eliminates the need to take the preferences or strategies of other agents into account.
Strategyproofness thus yields a robust prediction of equilibrium behavior. 
These and other advantages explain the popularity of strategyproofness as an incentive concept
\citep{PathakSoenmez2008LevelingThePlayingFieldSincereAndSophisticatedPlayersInTheBostonMechanism,AbdulkadirogluPathakRoth2009StrategyProofnessVersusEfficiencyInMatchingWithIndifferencesRedesigningTheNYCHighSchoolMatch}. 

The advantages of strategyproofness, however, come at a cost: 
\cite{Zhou1990OnAConjectureByGaleAboutOneSidedMatchingProblems} showed that, in the assignment problem, it is impossible to achieve the optimum with respect to incentives, efficiency, and fairness simultaneously.%
\footnote{Specifically, \citet{Zhou1990OnAConjectureByGaleAboutOneSidedMatchingProblems} showed that no (possibly random and possibly cardinal) assignment mechanism can satisfy
strategyproofness,
ex-ante efficiency,
and symmetry.}
This makes the assignment problem an interesting mechanism design challenge.
For example, the random serial dictatorship mechanism is strategyproof and anonymous, but only ex-post efficient. 
The more demanding ordinal efficiency is achieved by the probabilistic serial mechanism, but any mechanism that achieves ordinal efficiency and symmetry cannot be strategyproof \citep{BogomolnaiaMoulin2001ANewSolutionToTheRandomAssignmentProblem}. 
Finally, rank efficiency, an even stronger efficiency concept, can be achieved with rank value mechanisms.%
\footnote{\emph{Rank efficiency} \citep{Featherstone2020RankEfficiencyModelingACommonPolicymakerObjective} requires that an assignment's rank distribution 
(i.e., the number of assigned first, second, etc. choices) 
cannot be first-order stochastically dominated by the rank distribution of any other feasible assignment.
This is strictly more demanding than ordinal efficiency.}
However, rank efficiency is incompatible with strategyproofness, even without additional fairness requirements \citep{Featherstone2020RankEfficiencyModelingACommonPolicymakerObjective}.
Obviously, strategyproofness is in conflict with many other desiderata, and mechanism designers are therefore interested in studying non-strategyproof mechanisms.
This highlights the need for good tools to capture the incentive properties of non-strategyproof mechanisms and to analyze what trade-offs are possible between incentives for truthtelling and other
desirable properties.

In practice, non-strategyproof mechanisms are ubiquitous.
Examples include variants of the Boston mechanism that are used in many cities for the assignment of seats at public schools,%
\footnote{Variants of the non-strategyproof Boston mechanism are used in
Minneapolis, Seattle, Lee County \citep{KojimaUenver2014TheBostonSchoolChoiceMechanismsAnAxiomaticApproach},
San Diego, Freiburg (Germany) \citep{DurMennleSeuken2018FirstChoiceMaximalAndFirstChoiceStableSchoolChoiceMechanisms},
and throughout the German state of Nordrhein Westfalen \citep{BastekHuesmannNax2015MatchingPracticesForSecondarySchoolsGermanyMiPCountryProfile21}.}
a rank-efficient mechanism for the assignment of teachers to positions through the Teach for America program \citep{Featherstone2020RankEfficiencyModelingACommonPolicymakerObjective},
a mechanism for the assignment of on-campus housing at MIT that minimizes the number of unassigned units,%
\footnote{Source: MIT Division of Student Life, retrieved September 8, 2015: https://housing.mit.edu/, and personal communication.}
and the HBS draft mechanism for the assignment of course schedules at Harvard Business School \citep{BudishCantillon2012TheMultiUniAssignmentProblemTheoryAndEvidenceFromCourseAllocationAtHarvard}.
It is therefore important to understand the incentive properties of these mechanisms beyond the fact that they are ``not fully strategyproof.''

The incompatibility of strategyproofness with other desiderata in theory and
the prevalence of non-strategyproof assignment mechanisms in practice
explain why researchers have been calling for useful relaxations of strategyproofness \citep{Budish2012MatchingVersusMechanismDesign}.%
\footnote{Prior work has already led to some useful concepts for this purpose, e.g., the comparison by \emph{vulnerability to manipulation} \citep{PathakSoenmez2013SchoolAdmissionsReformInChicagoAndEnglandComparingMechanismsByTheirVulnerabilityToManipulation}
or \emph{strategyproofness in the large} \citep{AzevedoBudish2019StrategyProofnessInTheLarge}. 
We describe their relation to partial strategyproofness in Appendix~\ref{APP:CONSISTENCY_VULNERABILITY_DOSP} and Section~\ref{SEC:PSP:RELATION}.}
In this paper, we introduce \emph{partial strategyproofness}, a new, relaxed notion of strategyproofness, which is particularly suited to the analysis of assignment mechanisms.
Next, we illustrate the definition of partial strategyproofness with a motivating example,
and then we explain our main results and what they mean for mechanism design.

\medskip
Consider a setting with three agents, conveniently named $1$, $2$, $3$, and three objects, $a$, $b$, $c$, with unit capacity.
Suppose that the agents' preferences are
\begin{eqnarray}
    P_1 & : & a \succ b \succ c, \\
    P_2 & : & b \succ a \succ c, \\
    P_3 & : & b \succ c \succ a,
\end{eqnarray}
and that 
the non-strategyproof probabilistic serial mechanism
\citep{BogomolnaiaMoulin2001ANewSolutionToTheRandomAssignmentProblem}
is used to assign the objects.
If 
all
agents report 
truthfully, then agent 1 receives $a$, $b$, $c$ with probabilities $3/4$, $0$, $1/4$ respectively.
If agent 1 instead reports
\begin{eqnarray}
    P_1' & : & b \succ a \succ c,
\end{eqnarray}
then these probabilities change to $1/2$, $1/3$, $1/6$.
Observe that whether or not the misreport $P_1'$ increases agent 1's expected utility depends on how intensely it prefers $a$ over $b$:
If $u_1(a)$ is close to $u_1(b)$, then agent 1 would benefit from the misreport $P_1'$.
If $u_1(a)$ is significantly larger than $u_1(b)$, then agent 1 would prefer to report truthfully.
Specifically, agent 1 prefers to report truthfully if $ 3/4 u_1(a) \geq u_1(b)$ (assuming $u_1(c)=0$).

Our definition of partial strategyproofness generalizes the intuition from this motivating example:
For some bound $r$ in $[0,1]$, we say that agent $i$'s utility function $u_i$ satisfies \emph{uniformly relatively bounded indifference with respect to $r$ (\URBIr)} if $r u_i(a) \geq u_i(b)$ holds whenever $i$ prefers $a$ to $b$ (after appropriate normalization).
Fixing a setting (i.e., the number of agents and objects and their capacities), 
a mechanism is \emph{$r$-partially strategyproof (in that setting)} if it makes truthful reporting a dominant strategy for any agent whose utility function satisfies \URBIr. 
For example, as we show in Section~\ref{SEC:APPLICATIONS:PS}, the probabilistic serial mechanism has a \emph{degree of strategyproofness} of exactly $r=3/4$ in the setting of the motivating example.
Thus, from a market design perspective, we can now give honest and useful strategic advice to agents facing this mechanism in this setting: 
They are best off reporting their preferences truthfully as long as they
value their second choice at least a factor $3/4$ less than their first choice. 

\medskip
We argue that partial strategyproofness is a natural and useful way to think about
the incentive properties of non-strategyproof assignment mechanisms,
and we present two main arguments that support this claim:

\textbf{Partial strategyproofness has a compelling axiomatic motivation (Sections~\ref{SEC:SP} and \ref{SEC:PSP:DECOMP}).}
We first prove that full strategyproofness can be decomposed into three simple axioms.
Each of the axioms restricts the way in which a mechanism can react when an agent swaps two consecutively ranked objects in its preference report (e.g., from $P_i:a\succ b$ to $P_i':b\succ a$).
\enlargethispage{1.5em}
\begin{enumerate}
\itemsep 0pt
    \item A mechanism is \emph{swap monotonic} if either the swap makes it more likely that the agent receives $b$ (the object that it claims to prefer), or the mechanism does not 
        change the agent's assignment at all.
    \item A mechanism is \emph{upper invariant} if, by swapping $a$ and $b$, the agent 
        cannot affect its chances for objects that it prefers to $a$. 
    \item A mechanism is \emph{lower invariant} if, by swapping $a$ and $b$, the agent cannot affect its chances for objects that it likes less than $b$.
\end{enumerate}
In words, swap monotonicity ensures that the mechanism responds to changes in the agent's preference reports by increasing the assignment for the object that the agent claims to prefer,
upper invariance essentially means that the agent cannot benefit from truncating its preference list \citep{Hashimoto2014TwoAxiomaticApproachesToTheProbabilisticSerialMechanism},
and lower invariance is the complement of upper invariance but for less-preferred objects. 
For our first main result, we show that strategyproofness can be decomposed into these three axioms:
A mechanism is strategyproof if and only if it is swap monotonic, upper invariant, and lower invariant (Theorem~\ref{THM:SP}).
Intuitively, lower invariance is the least important of the three axioms (see Remark~\ref{REM:DROPPING_AXIOMS} for two formal arguments that support this intuition),
and by dropping it, we arrive at the larger class of partially strategyproof mechanisms:
We show that in a given setting, a mechanism is $r$-partially strategyproof for some $r>0$ if and only if it is swap monotonic and upper invariant (Theorem~\ref{THM:PSP}).
Thus, partial strategyproofness describes the incentive properties of (non-strategyproof) mechanisms that satisfy the first two of the three axioms.

\textbf{Partial strategyproofness yields a parametric measure for ``how strategyproof'' an assignment mechanism is (Sections~\ref{SEC:PSP:MAXIMAL} and \ref{SEC:PSP:DOSP}).}
To quantify the extent to which a mechanism makes truthful reporting a dominant strategy,
we consider the maximal value of $r$ for which the mechanism is $r$-partially strategyproof.
We call this value the \emph{degree of strategyproofness}.
Since $r$-partial strategyproofness is equivalent to strategyproofness on the restricted domain of utility functions that satisfy \URBIr,
a lower degree of strategyproofness corresponds to strategyproofness on a smaller restricted domain.
We prove maximality of the domain restriction by showing that
a utility function satisfies \URBIr\
if and only if
truthful reporting is a dominant strategy under \emph{all} $r$-partially strategyproof mechanisms for an agent with that utility function (Proposition~\ref{PROP:MAXIMALITY}).
Thus, without additional information about a mechanism
besides the fact that it is $r$-partially strategyproof, 
\URBIr\ is the only set of utility functions for which truthful reporting is guaranteed to be a dominant strategy.
In this sense, the degree of strategyproofness is a tight measure for incentive properties.

The degree of strategyproofness parametrizes a \textit{spectrum of incentive properties}: Since all utility functions satisfy $\text{URBI}(1)$, full strategyproofness is obviously the upper limit concept for $r$-partial strategyproofness as $r$ approaches 1.
Regarding the lower limit, \emph{lexicographic dominance} (\emph{LD}) \citep{Cho2018ProabilisticAssignmentAnExtensionApproach} is the weakest among the common dominance notions,%
\footnote{Lexicographic dominance is implied by 
\emph{trivial}, 
\emph{deterministic}, 
\emph{sure thing}, 
and \emph{first-order stochastic} dominance, 
but it is the only one among them under which all pairs of lotteries over objects are comparable \citep{AzizBrandBrill2013OnTheTradeoffBetweenEconomicEfficiencyAndStrategyProofnessInRandomizedSocialChoice}.} 
and the corresponding \emph{LD-strategyproofness} is therefore a minimal notion of strategyproofness for random assignment mechanisms. 
We prove that LD-strategyproofness is the lower limit concept for $r$-partial strategyproofness as $r$ approaches 0 (Proposition~\ref{PROP:SPECTRUM_LIMITS}).
The degree of strategyproofness thus parametrizes the entire spectrum 
between full strategyproofness and LD-strategyproofness.

\medskip
To complete the picture, we establish the formal relationships between partial strategyproofness and other relaxed notions of strategyproofness from prior work (Section~\ref{SEC:PSP:RELATION}).
For the assignment problem,
\citet{BogomolnaiaMoulin2001ANewSolutionToTheRandomAssignmentProblem} used \emph{weak SD-strategyproofness}
to describe the incentive properties of the probabilistic serial mechanism,
and \citet{Balbuzanov2016ConvexStrategyproofnessWithAnApplicationToTheProbabilisticSerialMechanism}
used \emph{convex strategyproofness} to refine their results.
We show that partial strategyproofness implies both notions but is not implied by either.
\emph{Approximate strategyproofness} \citep{Carroll2013AQuantitativeApproachToIncentivesApplicationToVotingRules}
and
\emph{strategyproof in the large} \citep{AzevedoBudish2019StrategyProofnessInTheLarge} are two additional notions that are useful in a wider range of mechanism design problems
beyond the assignment problem.  
When focusing on the assignment problem, 
we show that partial strategyproofness implies approximate strategyproofness in a meaningful way
and
that convergence of a mechanism's degree of strategyproofness to 1 in large markets implies strategyproofness in the large. Thus, establishing that a mechanism is partially strategyproof immediately provides additional insights about its incentive properties 
in terms of these notions from prior work. 

To illustrate the usefulness of the partial strategyproofness concept in mechanism design, we provide two applications (Section~\ref{SEC:APPLICATIONS}):
First, it enables the strongest description 
of the incentive properties of the probabilistic serial mechanism 
in finite settings 
known to date
(refining \citep{BogomolnaiaMoulin2001ANewSolutionToTheRandomAssignmentProblem,Balbuzanov2016ConvexStrategyproofnessWithAnApplicationToTheProbabilisticSerialMechanism})
as well as in the limit as markets get large (refining \citep{KojimaManea2010IncentivesInTheProbabilisticSerialMechanism,AzevedoBudish2019StrategyProofnessInTheLarge}).
Second, partial strategyproofness yields a distinction between two common variants of the Boston mechanism for school choice: 
Under the classic \emph{Boston mechanism (BM)} \citep{AbdulkadirogluSoenmez2003SchoolChoiceAMechanismDesignApproach},
all applications in round $k$ go to the students' $k^{\text{th}}$ choices,
but under the \emph{adaptive Boston mechanism (ABM)} \citep{Alcalde1996ImplemetationOfStableSolutionsToMarriageProblems,Miralles2008TheCaseForTheBostonMechanism,Harless2019ImmediateAcceptanceWithOrWithoutSkipsComparingSchoolAssignmentProcedures,Dur2015TheModifiedBostonMechanism}, students apply to their respective \emph{best available} school in each round.
Intuitively, one would expect ABM to have better incentive properties than BM because strategic skipping of exhausted schools is not a useful heuristic for manipulation under ABM.
However, formalizing this intuition is surprisingly challenging \citep{DurMennleSeuken2018FirstChoiceMaximalAndFirstChoiceStableSchoolChoiceMechanisms}.
Using our new partial strategyproofness concept, we can now show formally that ABM indeed has better incentive properties than BM:
When priorities are sufficiently random,%
\footnote{Concretely, \emph{sufficiently random} means that priorities must support all single priority profiles. In Remark~\ref{REM:COARSE_PRIORITIES}, 
we discuss 
an approach that allows an extension of partial strategyproofness
to problems with
coarse priorities.} ABM is partially strategyproof while BM is not.

Partial strategyproofness
is relevant for both theory and practical applications beyond the two concrete examples discussed above.
For theory, the simplicity of our axioms and our decomposition results enable new approaches for axiomatic analysis.%
\footnote{For example, subsequent work has already used our results to explore alternative combinations of the axioms
\citep{ChunYun2019UpperContourStrategyProofnessInTheProbabilisticAssignmentProblem,%
Noda2019APlannerOptimalMatchingMechanismAndItsIncentiveCompatibilityInARestrictedDomain,%
MennleSeuken2017TwoNewImpossibilityResultsForRandomTheAssignmentProblem}, to
identify when truthful reporting is a strictly dominant strategy under the deferred acceptance mechanism \citep{FackGernetHe2015WPBeyondTruthtellingPreferenceEstimationWithCentralizedSchoolChoice},
and to study ordinal Bayesian incentive compatibility in the random assignment problem \citep{DasguptaMishra2020OrdinalBayesianIncentiveCompatibilityInRandomAssignmentModel}.}
Furthermore, the parametric nature of partial strategyproofness enables a new quantitative analysis of the possible and necessary trade-offs between strategyproofness and other mechanism design objectives
(see, e.g., \citep{MennleSeuken2017HybridMechanismsTradingOffStrategyproofnessAndEfficiencyOfRandomAssignmentMechanisms}).
For practical applications, partial strategyproofness allows policy makers (e.g., school boards) to give honest and useful strategic advice to participating agents: 
An agent is best off reporting truthfully if its preference intensities differ sufficiently between any two objects; 
otherwise, the agent's potential gain from misreporting may be positive but it is quantifiably bounded (because of the formal relationship between partial and approximate strategyproofness).
Furthermore, partial strategyproofness can be helpful to assess the extent to which misreporting may be a concern 
(e.g., by enabling mechanism designers to estimate for how many agents 
misreporting is potentially beneficial). 
Therefore, we expect partial strategyproofness to become  a useful addition to the mechanism designer's toolbox for the analysis of non-strategyproof assignment mechanisms.
\section{Preliminaries}
\label{SEC:PRELIMINARIES}
\subsection{Model}
\label{SEC:PRELIMINARIES:MODEL}
A \emph{setting} $(N,M,q)$ consists of
a set of \emph{agents} $N$ with $n=|N|$,
a set of \emph{objects} $M$ with $m=|M|\geq 2$,
and \emph{capacities} $q = (q_1,\ldots,q_m)$
(i.e., $q_j$ units of object $j$ are available).
We assume $n \leq \sum_{j \in M} q_j$ (i.e., there are not more agents than the total number of units;
otherwise we include a dummy object with capacity $n$).
Each agent $i\in N$ has a strict \emph{preference order} $P_i$ over objects, where $P_i : a \succ b$ indicates that agent $i$ prefers object $a$ to object $b$.
Let $\mathcal{P}$ be the set of all possible preference orders over $M$.
A \emph{preference profile} $P = (P_i)_{i \in N}\in \mathcal{P}^N$ is a collection of preference orders of all agents, and $P_{-i} \in \mathcal{P}^{N\backslash\{i\}}$ is a collection of preference orders of all agents except $i$.
We assume that agents have von Neumann--Morgenstern \emph{utility functions}, where
agent $i$'s utility function $u_i : M \rightarrow \mathds{R}^+$ is 
\emph{consistent with $P_i$} (i.e., $u_i(a) > u_i(b)$ whenever $P_i : a \succ b$), and $U_{P_i}$ denotes the set of all utility functions that are consistent with $P_i$.

An \emph{assignment} is represented by an $n \times m$-matrix $x = (x_{i,j})_{i \in N, j\in M}$.
The value $x_{i,j}$ is the probability that agent $i$ gets object $j$.
The $i^{\text{th}}$ row $x_i = (x_{i,j})_{j\in M}$ of $x$ is called the \emph{assignment vector of $i$} (or \emph{$i$'s assignment}).
$x$ is \emph{feasible} if no object is assigned beyond capacity (i.e., $\sum_{i \in N} x_{i,j} \leq q_j$ for all $j \in M$) and each agent receives some object with certainty (i.e., $\sum_{j \in M} x_{i,j} = 1$ for all $i \in N$ and $x_{i,j} \geq 0$ for all $i \in N, j \in M$).
Unless explicitly stated otherwise, we only consider feasible assignments throughout this paper.
$x$ is called \emph{deterministic} if $x_{i,j} \in \{0,1\}$ for all $i\in N, j \in M$.
The Birkhoff--von Neumann Theorem and its extensions \citep{BudishCheKojimaMilgrom2013DesignRandomAllocMechsTheoryAndApplications} ensure that, for any given assignment, we can find a lottery over deterministic assignments that implements the respective marginal probabilities.
Finally, we denote by $X$ and $\Delta(X)$ the sets of all deterministic and random assignments, respectively.

A \emph{mechanism} is a mapping $ \f : \mathcal{P}^N \rightarrow \Delta(X)$ that selects an assignment based on a preference profile.
$\f_i(P_i,P_{-i})$ denotes $i$'s assignment when $i$ reports $P_i$ and the other agents report $P_{-i}$.
The mechanism $\f$ is \emph{deterministic} if it only selects deterministic assignments (i.e., $\f : \mathcal{P}^N \rightarrow X$).
Finally, given an agent $i$ with a utility function $u_i$ who reports $P_i$ when the other agents report $P_{-i}$, that agent's \emph{expected utility} is
$\mathds{E}_{\f_i(P_i,P_{-i})}[u_i] = \sum_{j \in M} u_i(j) \f_{i,j}(P_i,P_{-i})$.
\subsection{Strategyproofness and Auxiliary Concepts}
\label{SEC:PRELIMINARIES:CONCEPTS}
The most common incentive concept for the design of assignment mechanisms requires that agents maximize their expected utility by reporting their ordinal preferences truthfully, independent of the other agents' preference reports.
\begin{definition}
A mechanism $\f$ is \emph{expected-utility strategyproof (EU-strategyproof)} if,
for all agents $i \in N$,
all preference profiles $(P_i,P_{-i}) \in \mathcal{P}^N$,
all misreports $P_i' \in \mathcal{P}$,
and all utility functions $u_i \in U_{P_i}$,
we have $\mathds{E}_{\f_i(P_i,P_{-i})}[u_i] \geq \mathds{E}_{\f_i(P_i',P_{-i})}[u_i]$.
\end{definition}
Alternatively, strategyproofness can be defined in terms of stochastic dominance:
For a preference order $P_i \in \mathcal{P}$ and two assignment vectors $x_i$ and $y_i$,
we say that \emph{$x_i$ stochastically dominates $y_i$ at $P_i$}
if, for all objects $a \in M$, we have
\begin{equation}
    \sum_{j\in M \text{ with } P_i:j \succ a} x_{i,j} \geq \sum_{j\in M \text{ with } P_i:j \succ a} y_{i,j}.
    \label{EQ:SD_CONDITION}
\end{equation}
In words, $i$'s chances of obtaining one of its top-$k$ choices are weakly higher under $x_i$ than under $y_i$, for all ranks $k$. 
A mechanism $\f$ is \emph{stochastic dominance strategyproof (SD-strategyproof)} if,
for all agents $i \in N$,
all preference profiles $(P_i,P_{-i}) \in \mathcal{P}^N$,
and all misreports $P_i' \in \mathcal{P}$,
$i$'s assignment $\f_i(P_i,P_{-i})$ stochastically dominates $\f_i(P_i',P_{-i})$ at $P_i$.
SD-strategyproofness and EU-strategyproofness are equivalent \citep{Erdil2014StrategyProofStochasticAssignment}, and we therefore refer to this property as \emph{(full) strategyproofness}. 

\citet{Cho2016IncentivePropertiesForOrdinalMechanisms}
presented a substantially weaker incentive concept: 
Agents with lexicographic preferences 
prefer any (arbitrarily small) increase in the probability for a more-preferred object 
to any (arbitrarily large) increase in the probability for any less-preferred object;
and the associated lexicographic dominance strategyproofness requires that such agents 
prefer to report their preferences truthfully. 
\begin{definition}
\label{DEF:LD}
For a preference order $P_i\in \mathcal{P}$ and assignment vectors $x_i, y_i$,
we say that
\emph{$x_i$ lexicographically dominates} \emph{$y_i$ at $P_i$} if
$x_i=y_i$,
or
$x_{i,a}>y_{i,a}$ for some $a \in M$ and $x_{i,j} = y_{i,j}$ for all $j \in U(a,P_i)$.
A mechanism $\f$ is \emph{lexicographic dominance strategyproof} (\emph{LD-strategyproof}) if,
for
all agents $i \in N$,
all preference profiles $(P_i,P_{-i}) \in \mathcal{P}^N$,
and all misreports $P_i' \in \mathcal{P}$,
$\f_i(P_i,P_{-i})$ lexicographically dominates $\f_i(P_i',P_{-i})$ at $P_i$.
\end{definition}

Finally, we introduce the auxiliary concepts of neighborhoods and contour sets:
The \emph{neighborhood of a preference order $P_i$} is the set of all preference orders that differ from $P_i$ by a swap of two consecutively ranked objects, denoted $N_{P_i}$ (e.g.,
the neighborhood of $P_i: a \succ b\succ c \succ d$ contains $P_i' : b \succ a \succ c  \succ d$ but does not contain $P_i'': c\succ a \succ b \succ d$).
The \emph{upper contour set of $j$ at $P_i$} is the set of objects that $i$ strictly prefers to $j$, denoted $U(j,P_i)$.
Conversely, the \emph{lower contour set of $j$ at $P_i$} is the set of objects that $i$ likes strictly less than $j$, denoted $L(j,P_i)$.
For example, the upper contour set of $b$ at $P_i: a\succ b\succ c\succ d$ is $U(b,P_i)=\{a\}$ and the lower contour set is $L(b,P_i)=\{c,d\}$. 
\section{A New Decomposition of Strategyproofness}
\label{SEC:SP}
In this section, we introduce three axioms and we show that strategyproofness decomposes into these axioms. 
Swapping two consecutively ranked objects in the true preference order 
(or equivalently, reporting a preference order from its neighborhood) 
is a basic kind of misreport.
The axioms we define limit the ways in which a mechanism can change an agent's assignment when that agent uses such a basic misreport.
\begin{axiom}[Swap Monotonicity]
\label{AX:SM}
A mechanism $\f$ is \emph{swap monotonic} if,
for
all agents $i \in N$,
all preference profiles $(P_i,P_{-i}) \in \mathcal{P}^N$,
and all misreports $P_i' \in N_{P_i}$
with $P_i : a \succ b$ but $P_i': b \succ a$,
one of the following holds:
\begin{itemize}
\setlength{\itemsep}{0pt}
    \item either: $\f_i(P_i,P_{-i}) = \f_i(P_i',P_{-i})$,
    \item or: $\f_{i,b}(P_i',P_{-i}) > \f_{i,b}(P_i,P_{-i})$.
\end{itemize}
\end{axiom}
In words, swap monotonicity requires that the mechanism reacts to the swap in a \emph{direct} and \emph{monotonic} way:
If the swap that brings $b$ forward affects the agent's assignment at all, then at least its assignment for $b$ must be affected directly.
Moreover, this change must be monotonic in the sense that the agent's assignment for $b$ must increase 
when $b$ is reportedly more preferred.

For deterministic mechanisms, swap monotonicity is equivalent to strategyproofness
(Proposition~\ref{PROP:DET_MECH_SM_SP_EQUIV} in Appendix~\ref{APP:DET_MECH_SM_SP_EQUIV}).
For the more general class of random mechanisms, swap monotonicity is weaker than strategyproofness but prevents a certain obvious kind of manipulability:
Consider a mechanism that assigns an agent's reported first choice with probability $1/3$ and its reported second choice with probability $2/3$.
The agent is unambiguously better off by ranking its second choice first.
Swap monotonicity precludes such opportunities for manipulation.
Nevertheless, even swap monotonic mechanisms may be manipulable in a stochastic dominance sense, as the next example shows.
\begin{example}
\label{EX:MECH_SM_FOSD_MANIP}
Consider a setting with four objects $a,b,c,d$ and a single agent $i$ with preference order $P_i:a\succ b\succ c \succ d$.
Suppose that reporting a preference for $b$ over $c$
leads to an assignment of $(0,1/2,0,1/2)$ for $a,b,c,d$ respectively,
and reporting a preference for $c$ over $b$
leads to $(1/2,0,1/2,0)$.
This is consistent with swap monotonicity;
yet, the latter assignment stochastically dominates the former at $P_i$.
\end{example}
Note that the misreport in Example~\ref{EX:MECH_SM_FOSD_MANIP} affects the agent's assignment for $a$, an object that the agent strictly prefers to both $b$ and $c$, the objects that get swapped.
Our next axiom precludes such effects.
\begin{axiom}[Upper Invariance]
\label{AX:UI}
A mechanism $\f$ is \emph{upper invariant} if, for
all agents $i \in N$,
all preference profiles $(P_i,P_{-i}) \in \mathcal{P}^N$,
and all misreports $P_i' \in N_{P_i}$
with $P_i : a \succ b$ but $P_i': b \succ a$,
we have that $\f_{i,j}(P_i,P_{-i}) = \f_{i,j}(P_i',P_{-i})$ for all $j \in U(a,P_i)$.
\end{axiom}
Intuitively, upper invariance ensures that agents cannot influence their chances of obtaining more-preferred objects by changing the order of less-preferred objects.
The axiom was originally introduced by \citet{Hashimoto2014TwoAxiomaticApproachesToTheProbabilisticSerialMechanism} (who called it \emph{weak invariance}).
It was one of the central axioms in their characterization of the probabilistic serial mechanism.
If an outside option is available and if the mechanism is individually rational, then upper invariance is equivalent to \emph{truncation
robustness} (i.e., no agent can benefit by ranking the outside option above acceptable objects).
Many assignment mechanisms satisfy upper invariance,
including
random serial dictatorship,
probabilistic serial,
the Boston mechanism,
deferred acceptance (for agents on the proposing side),
top-trade cycles,
and the HBS draft mechanism.

Finally, our third axiom is symmetric to upper invariance but restricts how swaps can affect the assignment for less-preferred objects.
\begin{axiom}[Lower Invariance]
\label{AX:LI}
A mechanism $\f$ is \emph{lower invariant} if, for
all agents $i \in N$,
all preference profiles $(P_i,P_{-i}) \in \mathcal{P}^N$,
and all misreports $P_i' \in N_{P_i}$
with $P_i : a \succ b$ but $P_i': b \succ a$,
we have that $\f_{i,j}(P_i,P_{-i}) = \f_{i,j}(P_i',P_{-i})$ for all $j \in L(b,P_i)$.
\end{axiom}
In words, a mechanism is lower invariant if changing the order of two consecutively ranked objects does not affect the agents' chances of obtaining any less-preferred objects.
Examples of lower invariant mechanisms are strategyproof mechanisms, like
random serial dictatorship and
top-trade cycles. 
constrained serial dictatorship  \citep{Noda2019APlannerOptimalMatchingMechanismAndItsIncentiveCompatibilityInARestrictedDomain} is swap monotonic and lower invariant, 
and the Deferred Acceptance is lower invariant for agents on both sides. 

Each of the three axioms affects incentives by preventing misreports from being beneficial in particular ways:
swap monotonicity forces mechanisms to change the assignment for the respective objects directly and in the right direction,
upper invariance is essentially equivalent to truncation robustness,
and lower invariance mirrors upper invariance but for less-preferred objects.
In combination, they give rise to our first main result, the decomposition of strategyproofness into these axioms.
\begin{theorem}[Decomposition of Strategyproofness]
\label{THM:SP}
A mechanism is strategyproof
if and only if
it is swap monotonic, upper invariant, and lower invariant.
\end{theorem}
\begin{proof}
    \emph{Sufficiency ($\Rightarrow$).}
    Assume towards contradiction that a mechanism $\f$ is strategyproof but not upper invariant.
    Then there exists an agent $i\in N$, a preference profile $P = (P_i,P_{-i}) \in \mathcal{P}^N$, a misreport $P_i' \in N_{P_i}$ with $P_i:a \succ b$ but $P_i':b \succ a$, and an object $j$ which $i$ prefers strictly to $a$ with $\f_{i,j}(P_i',P_{-i}) \neq \f_{i,j}(P_i,P_{-i})$.
    Without loss of generality, let $j$ be $i$'s most-preferred such object
    and $\f_{i,j}(P_i',P_{-i}) > \f_{i,j}(P_i,P_{-i})$ (otherwise, reverse the roles of $P_i'$ and $P_i$).
    Then $\f_i(P_i,P_{-i})$ does not stochastically dominate $\f_j(P_i',P_{-i})$,
    a contradiction to SD-strategyproofness.
    Lower invariance follows analogously, except that we take $j$ to be the \emph{least}-preferred object for which $i$'s assignment changes.

    By upper and lower invariance, 
    any swap of consecutively ranked objects 
    (e.g., from $P_i:a \succ b$ to $P_i':b \succ a$) 
    leads to a re-distribution of probability between $a$ and $b$.
    If reporting $P_i'$ leads to a strictly higher assignment for $a$, then $\f_i(P_i',P_{-i})$ strictly stochastically dominates $\f_i(P_i,P_{-i})$ at $P_i$, a contradiction.
    This implies swap monotonicity.

    \medskip
    \emph{Necessity ($\Leftarrow$).}
    We invoke a result of \citet{Carroll2012WhenAreLocalIncentiveConstraintsSufficient} that strategyproofness can be shown by verifying that no agent can benefit from swapping two consecutively ranked objects.
    Let $\f$ be a swap monotonic, upper invariant, and lower invariant mechanism, and consider an agent $i\in N$, a preference profile $(P_i,P_{-i}) \in \mathcal{P}^N$, and a misreport $P_i' \in N_{P_i}$ with $P_i:a \succ b$ but $P_i':b \succ a$.
    Observe that $\f_{i}(P_i,P_{-i})$ stochastically dominates $\f_{i}(P_i',P_{-i})$ at $P_i$:
    By upper and lower invariance, $i$'s assignment for all objects remains constant under the misreport, except possibly for $a$ and $b$;
    and, by swap monotonicity, $i$'s assignments for $a$ and $b$ can only decrease and increase, respectively.
\end{proof}
Theorem~\ref{THM:SP} highlights that strategyproofness is quite restrictive:
If an agent swaps two objects (e.g., from $P_i:a \succ b$ to $P_i': b \succ a$), the only way in which a strategyproof mechanism can react is by increasing that agent's assignment for $b$ and decreasing its assignment for $a$ by the same amount.
The appeal of our decomposition in Theorem~\ref{THM:SP} lies in the choice of axioms, which are simple and easily motivated.
They make the decomposition particularly useful,
for example when verifying strategyproofness of mechanisms
or when encoding strategyproofness as constraints under the automated mechanism design paradigm \citep{Sandholm2003AutomatedMechanismDesignANewApplicationAreaForSearchAlgorithms}.%
\footnote{The decomposition has been used in subsequent work, 
e.g., to identify when truthtelling is a strictly dominant strategy under the deferred acceptance mechanism \citep{FackGernetHe2015WPBeyondTruthtellingPreferenceEstimationWithCentralizedSchoolChoice}
and
to study Bayesian incentive compatibility of ordinal mechanisms \citep{DasguptaMishra2020OrdinalBayesianIncentiveCompatibilityInRandomAssignmentModel}.}
\begin{remark}
\label{RM:SP_INDIFF}
While we focus on strict preferences in the present paper, Theorem~\ref{THM:SP} can be extended to the case of weak preferences; see \citep{MennleSeuken2017AnAxiomaticDecompositionOfStrategyproofnessForOrdinalMechanismsWithIndifferences}.
\end{remark}
\section{Partial Strategyproofness}
\label{SEC:PSP}
Recall the motivating example from the introduction,
where agent 1 was contemplating a misreport under the probabilistic serial mechanism.
$r = 3/4$ was the pivotal ratio between the agent's (normalized) utility for its first and second choices
which determined whether the misreport was beneficial or not.
The following domain restriction generalizes the intuition from this basic example to agents whose preference intensities differ sufficiently between any two objects.
\begin{definition}
\label{DEF:URBI}
A utility function $u_i$ satisfies \emph{uniformly relatively bounded indifference with respect to $r \in [0,1]$ (\URBIr)} if, for all objects $a,b\in M$ with $u_i(a) > u_i(b)$,
\begin{equation}
    r \left( u_i(a) -\min_{j \in M} u_i(j) \right) \geq u_i(b) -\min_{j \in M} u_i(j).
\label{EQ:URBI_CONSTRAINT}
\end{equation}
\end{definition}
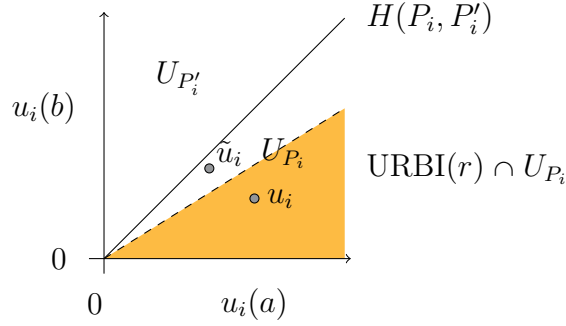
\begin{figure}%
\begin{center}
\begin{tikzpicture}[scale=4]
    \fill[fill=Dandelion] (0,0) -- (.8,0) -- (.8,.5);
    \draw[->] (-0.05,0) -- (0.82,0) coordinate (x axis);
    \draw[->] (0,-0.05) -- (0,0.82) coordinate (y axis);
    \draw[-,dashed] (0,0) -- (.8,.5);
    \draw (0, -0.15) node {$0$};
    \draw (-0.15,0) node {$0$};
    \draw (0.5,-0.15) node { $u_i(a)$};
    \draw (-0.2,.5) node { $u_i(b)$};
    \draw[-] (0,0) -- (.8,.8);
    \draw [fill=Gray](0.5,0.2) circle [radius=0.015];
    \node[label=right:{$u_i$}] at (.47,.2) {};
    \draw [fill=Gray](0.35,0.3) circle [radius=0.015];
    \node[label=right:{$\tilde{u}_i$}] at (.3,.35) {};
    \node[label=right:{$U_{P_i}$}] at (.45,.35) {};
    \node[label=right:{$U_{P'_i}$}] at (.1,.6) {};
    \node[label=right:{$H(P_i,P'_i)$}] at (.8,.8) {};
    \node[label=right:{$\URBIr\cap U_{P_i}$}] at (.8,.3) {};
\end{tikzpicture}%
\end{center}
\caption{Illustration of uniformly relatively bounded indifference.}%
\label{fig:geom_urbi}
\end{figure}
If $\min_{j\in M}\left(u_i(j)\right) = 0$ (i.e., $i$ has utility $0$ for its last choice), then Inequality (\ref{EQ:URBI_CONSTRAINT}) simplifies to
\begin{equation}
r u_i(a) \geq u_i(b).
\end{equation}
In words, \URBIr\ requires that agent $i$ values $b$ at least a factor $r$ less than $a$.
Consider Figure~\ref{fig:geom_urbi}, which depicts the utility space for a setting with only two objects $a$ and $b$, such that every utility function corresponds to a point in this figure. This  allows the following geometric interpretation: The \URBIr\ condition means that $i$'s utility function (e.g., the point labeled $u_i$) cannot be arbitrarily close to the indifference hyperplane $H(P_i,P'_i)$,
but it must lie within the shaded triangular area.
$r$ is the slope of the dashed line that bounds this area at the top.
Any utility function in $U_{P_i}$ that lies outside the shaded area (e.g., the point labeled $\tilde{u}_i$) violates \URBIr.
For convenience, we also use \URBIr\ to denote the \emph{set of utility functions} that satisfy the constraints.

We are now ready to formally define our new incentive concept.
\begin{definition}
\label{DEF:PSP}
Given a setting $(N,M,q)$ and a bound $r \in [0,1]$,
a mechanism $\f$ is \emph{$r$-partially strategyproof in $(N,M,q)$}
if, for
all agents $i \in N$,
all preference profiles $(P_i,P_{-i}) \in \mathcal{P}^N$,
all misreports $P_i' \in \mathcal{P}$,
and all utility functions $u_i \in U_{P_i} \cap \URBIr$,
we have $\mathds{E}_{\f_i(P_i,P_{-i})}[u_i] \geq \mathds{E}_{\f_i(P_i',P_{-i})}[u_i]$.
\end{definition}
In words, an $r$-partially strategyproof mechanism makes truthful reporting a dominant strategy at least for those agents whose utility functions satisfy \URBIr. 
Note that the value of $r$ depends on the setting. 
Since all utility functions satisfy $\text{URBI}(1)$, $1$-partial strategyproofness is equivalent to full strategyproofness, and since the sets \URBIr\
are nested (i.e., $\URBIr \subseteq \textrm{URBI}(r')$ for $r \leq r'$), 
the $r$-partial strategyproofness requirement is less demanding for smaller $r$.
\subsection{A Decomposition of Partial Strategyproofness}
\label{SEC:PSP:DECOMP}
For our second main result, we prove that dropping lower invariance but still requiring swap monotonicity and upper invariance leads to partial strategyproofness.
\begin{theorem}[Decomposition of Partial Strategyproofness]
\label{THM:PSP}
Given a setting $(N,M,q)$, a mechanism is $r$-partially strategyproof for some $r > 0$
if and only if it is swap monotonic and upper invariant.
\end{theorem}
\begin{proof}
Our three axioms swap monotonicity, upper invariance, and lower invariance
are \emph{local} constraints
in the sense that they only restrict how mechanisms can react to swaps. 
Conversely, full and partial strategyproofness are \emph{global} constraints
because they restrict reactions to arbitrary report changes. 
In our proof of Theorem~\ref{THM:SP}, 
we have used the local sufficiency result of \citet{Carroll2012WhenAreLocalIncentiveConstraintsSufficient} 
to bridge the gap between local and global constraints. 
To bridge the analogous gap in the proof of Theorem~\ref{THM:PSP}, 
we use a local sufficiency result for lexicographic dominance strategyproofness by \citet{Cho2016IncentivePropertiesForOrdinalMechanisms}. 

It is clear that swap monotonicity and upper invariance are equivalent to LD-\emph{adjacent} strategyproofness
(i.e., for
all $i \in N$,
$(P_i,P_{-i}) \in \mathcal{P}^N$,
and $P_i' \in N_{P_i}$,
$\f_i(P_i,P_{-i})$ lexicographically dominates $\f_i(P_i',P_{-i})$ at $P_i$).
By Theorem 1 of \citet{Cho2016IncentivePropertiesForOrdinalMechanisms},
swap monotonicity and upper invariance are thus equivalent to LD-strategyproofness.
With this, we must show the following:
\begin{lemma}
\label{LEM:PSP_LDSP_EQUIV}
    Given a setting $(N,M,q)$, a mechanism $\f$ is $r$-partially strategyproof for some $r>0$ if and only if $\f$ is LD-strategyproof.
\end{lemma}
We can restrict attention to mechanisms under which at least one agent has a direct influence on its own assignment;
formally, there exists an agent $i \in N$, a preference profile $(P_i,P_{-i})\in \mathcal{P}^N$, and a misreport $P_i' \in \mathcal{P}$ such that $\f_i(P_i,P_{-i}) \neq \f_i(P_i', P_{-i})$.%
\footnote{Mechanisms that violate this property are trivially strategyproof for all agents and thus both 
LD-strategyproof 
and 
$r$-partially strategyproof for all $r\in [0,1]$.}
Let $\delta$ be the smallest non-zero change in any agent's assignment resulting from any change of report by that agent, defined by
\begin{equation}
    \delta = \min \left\{
        \left|\f_{i,j}(P_i,P_{-i})-\f_{i,j}(P_i',P_{-i})\right|
        \left|
        \begin{array}{c}
            i \in N, j \in M, (P_i,P_{-i}) \in \mathcal{P}^N, P_i'\in \mathcal{P}, \\
            \text{s.t. }\left|\f_{i,j}(P_i,P_{-i})-\f_{i,j}(P_i',P_{-i})\right| > 0
        \end{array}
        \right.\right\}.
    \label{EQ:DEF_OF_DELTA_IN_PROOF_OF_THM_2}
\end{equation}
It is easy to see that $\delta$ is well-defined and strictly positive for the mechanisms we consider. 

\medskip
\emph{Necessity (partial strategyproofness $\Leftarrow$ LD-strategyproofness).}
For an agent $i\in N$ and a preference profile $(P_i,P_{-i}) \in \mathcal{P}^N$, suppose that $i$ considers misreport $P_i'\in \mathcal{P}$.
If $\f_i(P_i,P_{-i}) \neq \f_i(P_i',P_{-i})$, then LD-strategyproofness implies that there exists some object $a$ for which $i$'s assignment strictly decreases, and $i$'s assignment for all more-preferred objects $j \in U(a,P_i)$ remains unchanged.
Since $i$'s assignment for $a$ changes, it must decrease by at least $\delta$.
Let $b$ be the object that $i$ ranks directly below $a$ in $P_i$.
Then $i$'s gain in expected utility from misreporting is greatest if,
first, $i$ receives $a$ with probability $\delta$ and its last choice with probability $(1-\delta)$ when being truthful,
and second, $i$ receives $b$ with certainty when misreporting.
The gain is therefore bounded from above by
\begin{equation}
    u_i(b) - \left(\delta u_i(a) + (1-\delta) \min_{j\in M} u_i(j)\right),
\end{equation}
which is weakly negative if
\begin{equation}
    \delta \left(u_i(a) - \min_{j\in M} u_i(j)\right) \geq u_i(b) - \min_{j\in M} u_i(j).
\label{EQ:THM:PSP:GAIN_UPPER_BOUND}
\end{equation}
Inequality (\ref{EQ:THM:PSP:GAIN_UPPER_BOUND}) holds for all utility functions in $U_{P_i} \cap \URBI(\delta)$.
Thus, if $i$'s utility function satisfies $\URBI(\delta)$, then no misreport increases $i$'s expected utility,
or equivalently, $\f$ is $\delta$-partially strategyproof.

\medskip
\emph{Sufficiency (partial strategyproofness $\Rightarrow$ LD-strategyproofness).}
Let $\f$ be $r$-partially strategyproof for some $r>0$, and assume towards contradiction that $\f$ violates LD-strategyproofness.
This is equivalent to saying that there exists
an agent $i\in N$,
a preference profile $(P_i,P_{-i})\in \mathcal{P}^N$,
a misreport $P_i'\in \mathcal{P}$ with $\f_i(P_i,P_{-i}) \neq \f_i(P_i',P_{-i})$,
and an object $a \in M$,
such that
$\f_{i,a}(P_i,P_{-i}) < \f_{i,a}(P_i',P_{-i})$ but
$\f_{i,j}(P_i,P_{-i}) = \f_{i,j}(P_i',P_{-i})$ for all $j \in U(a,P_i)$.
Again, let $b$ be the object that $i$ ranks directly below $a$ in $P_i$.
Since $i$'s assignment for $a$ increases, it must increase by at least $\delta$.
Thus, $i$'s gain in expected utility is \emph{smallest} if,
first, $i$ receives $b$ with certainty when being truthful,
and second, $i$ receives $a$ with probability $\delta$ and its last choice with probability $(1-\delta)$ when misreporting.
\enlargethispage{1.5em}
This makes
\begin{equation}
    \left( \delta u_i(a) + (1-\delta) \min_{j\in M} u_i(j) \right) - u_i(b)
\end{equation}
a lower bound on $i$'s gain from misreporting.
This bound is strictly positive if
\begin{equation}
    \delta \left(u_i(a) - \min_{j\in M} u_i(j) \right) > u_i(b) - \min_{j\in M}u_i(j),
\end{equation}
which holds for all utility functions in $U_{P_i} \cap \URBIr$ for $r < \delta$.
Therefore, $\f$ cannot be $r$-partially strategyproof for any $r > 0$, a contradiction.
\end{proof}
Theorem~\ref{THM:PSP} provides an axiomatic motivation for our definition of partial strategyproofness:
The class of partially strategyproof mechanisms consists exactly of those mechanisms that are swap monotonic and upper invariant, but they may violate lower invariance.
\enlargethispage{1em}
\begin{remark}
\label{REM:DROPPING_AXIOMS}
Two main arguments suggest dropping lower invariance and keeping swap monotonicity and upper invariance as a sensible approach towards relaxing strategyproofness.
First, on the positive side,
upper invariance is essentially equivalent to robustness to manipulation by truncation \citep{Hashimoto2014TwoAxiomaticApproachesToTheProbabilisticSerialMechanism},
and,
for deterministic mechanisms, swap monotonicity is equivalent to strategyproofness (Proposition~\ref{PROP:DET_MECH_SM_SP_EQUIV} in Appendix~\ref{APP:DET_MECH_SM_SP_EQUIV}). 
Second, on the negative side, dropping either swap monotonicity or upper invariance (instead of lower invariance) does not admit the construction of interesting ordinally efficient mechanisms:
the probabilistic serial mechanism \citep{BogomolnaiaMoulin2001ANewSolutionToTheRandomAssignmentProblem} is swap monotonic, upper invariant (but not lower invariant), and ordinally efficient,
and it satisfies the additional axioms
\emph{symmetry},
\emph{anonymity},
\emph{neutrality},
and \emph{non-bossiness}.
In contrast,
no mechanism can be
upper invariant,
lower invariant,
ordinally efficient,
and symmetric;
and no mechanism can be
swap monotonic,
lower invariant,
ordinally efficient,
anonymous,
neutral,
and non-bossy.%
\footnote{See
\citep{MennleSeuken2017TwoNewImpossibilityResultsForRandomTheAssignmentProblem}
for formal definitions of these axioms
and both impossibility results.} 
This means that these popular combinations of mechanism design axioms become unattainable when dropping swap monotonicity or upper invariance (instead of lower invariance).

Despite these two arguments,
whether an axiom is more or less ``important'' is also a question of taste.
We have chosen to drop lower invariance in the present paper,
but investigating the consequences of dropping swap monotonicity or upper invariance is definitely an interesting 
research question.%
\footnote{In subsequent work,
\citet{Noda2019APlannerOptimalMatchingMechanismAndItsIncentiveCompatibilityInARestrictedDomain} considered the combination of swap monotonicity and lower invariance;
\citet{ChunYun2019UpperContourStrategyProofnessInTheProbabilisticAssignmentProblem} considered the combination of upper and lower invariance, 
and they independently proved the incompatibility with ordinal efficiency and symmetry.}
\end{remark}
\subsection{Maximality of the URBI(r) Domain Restriction}
\label{SEC:PSP:MAXIMAL}
In this section, we study how well the \URBIr\ domain restriction captures the incentive properties of non-strategyproof assignment mechanisms.
By definition, $r$-partial strategyproofness implies that the set \URBIr\ must be contained in the set of utility functions for which truthful reporting is a dominant strategy.
However, the two sets may not be exactly equal, as the following example shows.
\begin{example}
\label{EX:ABM_PSP_NON_URBI}
Consider a setting with 4 agents and 4 objects with unit capacity.
In this setting, the adaptive Boston mechanism with priorities determined by a single uniform lottery
($\ABM^{\mathds{U}}$, see Section~\ref{SEC:APPLICATIONS:BM})
is $1/3$-partially strategyproof but not $r$-partially strategyproof for any $r > 1/3$.
However, it is a simple (though tedious) exercise to verify that an agent with utility function
$u_i(a) = 6$, $u_i(b)=2$, $u_i(c)=1$, $u_i(d)=0$
cannot benefit from misreporting, independent of the reports from the other agents.
But $u_i$ violates $\URBI\left(1/3\right)$, since
$\left(u_i(c)-\min_{j\in M}u(j)\right)/\left(u_i(b)-\min_{j\in M}u(j)\right) = 1/2 > 1/3$.
Thus, the set of utility functions for which $\ABM^{\mathds{U}}$ makes truthful reporting a dominant strategy is strictly larger than the set $\text{URBI}(1/3)$.
\end{example}
Example~\ref{EX:ABM_PSP_NON_URBI} shows that, for some $r$-partially strategyproof mechanism, there may exist utility functions that violate \URBIr\ but for which truthful reporting is nonetheless a dominant strategy.
This raises the question whether \URBIr\ is ``too small'' in the sense that
it excludes some utility functions for which \emph{all} $r$-partially strategyproof mechanisms make truthful reporting a dominant strategy.
Our next proposition dispels this concern because it shows \emph{maximality} of the \URBIr\ domain restriction.
\begin{proposition}[Maximality]
\label{PROP:MAXIMALITY}
Consider a setting $(N,M,q)$ with $m\geq3$ objects,
a bound $r>0$,
and an agent $i$ with utility function $u_i$.
Then $u_i$ satisfies \URBIr\
if and only if
truthful reporting maximizes the expected utility for $i$ under all $r$-partially strategyproof, anonymous mechanisms. 
\end{proposition}
\begin{proof}
Towards contradiction, assume that a utility function $\tilde{u}_i$ violates \URBIr. 
Then
\begin{equation}
    \frac{\tilde{u}_i(b)-\min_{j \in M} \tilde{u}_i(j)}{\tilde{u}_i(a)-\min_{j \in M} \tilde{u}_i(j)} = \tilde{r}
    \label{EQ:VIOLATION_OF_URBI_BY_U_TILDE}
\end{equation}
for some $a,b\in M$ and $\tilde{r} > r$.
Observe that $b$ cannot be $i$'s last choice because $0/\left(\tilde{u}_i(a)-\min_{j \in M} \tilde{u}_i(j)\right) \leq r$ is trivially satisfied. 
We construct a mechanism $\tilde{\f}$ that is $r$-partially strategyproof but manipulable for agent $i$ with utility function $\tilde{u}_i$, 
setting the assignment for the distinguished agent $i$ as follows:
Fix parameters $\delta_a,\delta_b \in [0,1/m]$;
then,
for all $P_{-i} \in \mathcal{P}^N$,
all preference orders $\hat{P}_i \in \mathcal{P}$,
and all objects $j\in M$,
let
\begin{equation*}
    \tilde{\f}_{i,j}(\hat{P}_i,P_{-i}) = \left\{
        \begin{array}{ll}
            1/m, & \text{ if }\hat{P}_i:a\succ b, \\
            1/m
                + \delta_b \mathds{1}_{\{j=b\}}
             - \delta_a \mathds{1}_{\{j=a\}}
                + (\delta_a-\delta_b) \mathds{1}_{\{j=d\}}
                , & \text{ if }\hat{P}_i:b\succ a,
        \end{array}
    \right.
\end{equation*}
where $d$ is the last choice under $\hat{P}_i$.
For all other agents, distribute the remaining probabilities evenly.
With parameters $\delta_a, \delta_b \in[0,1/m]$, this mechanism is well-defined.
Next, let $\delta_b=1/m$ and choose $\delta_a \in [r/m,\tilde{r}/m)$. 
It is straightforward to verify that $\tilde{\f}$ has the desired properties (Lemma~\ref{LEM:VARPHI_TILDE_PSP_BUT_MANIP} in Appendix~\ref{APP:VARPHI_TILDE_PSP_BUT_MANIP_LEMMA}).
To obtain an anonymous mechanism with the same properties, we randomly assign each agent to the role of the distinguished agent $i$.
This yields a contradiction to the assumption that $i$ finds truthful reporting a dominant strategy under any $r$-partially strategyproof, anonymous mechanism. 

Sufficiency holds by definition of $r$-partial strategyproofness. 
\end{proof}
Proposition~\ref{PROP:MAXIMALITY} means that, for any utility function that violates \URBIr, there exists some $r$-partially strategyproof mechanism under which truthful reporting is not a dominant strategy for an agent with that utility function.
Thus, unless we are given additional structural information about the mechanism besides the fact that it is \mbox{$r$-partially} strategyproof (and possibly anonymous), \URBIr\ is in fact the largest set of utility functions for which truthful reporting is guaranteed to be a dominant strategy.
\subsection{The Degree of Strategyproofness}
\label{SEC:PSP:DOSP}
Partial strategyproofness induces a natural parameter to measure 
``how strategyproof'' a non-strategyproof assignment mechanism is. 
%
\begin{definition}
\label{DEF:DOSP}
Given a setting $(N,M,q)$ and a mechanism $\f$, we define the \emph{degree of strategyproofness of $\f$} (\emph{in the setting $(N,M,q)$}) as
\begin{equation}
    \rho_{(N,M,q)}(\f) = \max\left\{ r \in [0,1] ~|~ \f \text{ is }r\text{-partially strategyproof in }(N,M,q) \right\}.%
    \footnote{To see that $\rho_{(N,M,q)}$ is well-defined, observe that \URBIr\ is topologically closed.
    Thus, a mechanism that is $r'$-partially strategyproof for all $r' < r$ must also be $r$-partially strategyproof.}
\label{eq:def_rho}
\end{equation}
\end{definition}
Observe that, for $0 \leq r < r' \leq 1$ we have $\URBI(r) \subset \URBI(r')$ by construction.
Thus, a lower degree of strategyproofness corresponds to a weaker guarantee.
By maximality from Proposition~\ref{PROP:MAXIMALITY},
the degree of strategyproofness constitutes a meaningful measure for incentive properties:
If the only known attributes of $\f$ are that it is swap monotonic and upper invariant and that $r=\rho_{(N,M,q)}(\f)$,
then truthful reporting is guaranteed to be a dominant strategy for all agents with utility functions in $\text{URBI}(r)$,
but it is impossible to give such a guarantee for any utility function that violates $\text{URBI}(r)$. 

The degree of strategyproofness can be used to compare two mechanisms by their incentive properties:
$\rho_{(N,M,q)}(\f) > \rho_{(N,M,q)}(\g)$
means that $\f$ makes truthful reporting a dominant strategy on a strictly larger \URBIr\ domain restriction than $\g$ does.%
\footnote{\citet{PathakSoenmez2013SchoolAdmissionsReformInChicagoAndEnglandComparingMechanismsByTheirVulnerabilityToManipulation} proposed a concept to compare mechanisms by their \emph{vulnerability to manipulation}.
As Proposition~\ref{PROP:DOSP_ISM_CONSISTENT} in Appendix~\ref{APP:CONSISTENCY_VULNERABILITY_DOSP} shows, this comparison is consistent with (but not equivalent to) the comparison of mechanisms by their degrees of strategyproofness.}
In Section~\ref{SEC:APPLICATIONS}, we apply this comparison to differentiate three assignment mechanisms, namely the probabilistic serial mechanism and two variants of the Boston mechanism. 
Furthermore, in  \citep{MennleSeuken2017HybridMechanismsTradingOffStrategyproofnessAndEfficiencyOfRandomAssignmentMechanisms}
we have used the degree of strategyproofness 
to quantify trade-offs between strategyproofness and efficiency that are achievable via hybrid mechanisms (i.e., convex combinations of a strategyproof mechanism and a manipulable 
mechanism with better efficiency properties).

\medskip
Observe that the degree of strategyproofness parametrizes a spectrum of incentive concepts,
where the upper limit of this spectrum is full strategyproofness.
Conversely, we would also like to understand what the \emph{lower limit} is.
To answer this question, we consider the weaker notion of lexicographic dominance strategyproofness
(Definition \ref{DEF:LD}).
Recall that a mechanism is
LD-strategyproof if truthful reporting is a dominant strategy for agents who
prefer any (arbitrarily small) increase in the probability for a more-preferred object
to any (arbitrarily large) increase in the probability for any less-preferred object. 
The next proposition formalizes the upper and lower limits for the partial strategyproofness concept.
\enlargethispage{1em}
\begin{proposition}[Limit Concepts]
\label{PROP:SPECTRUM_LIMITS}
Given a setting $(N,M,q)$ and an $r>0$,
let $\text{SP}(N,M,q)$, $r\text{-PSP}(N,M,q)$, and $\text{LD-SP}(N,M,q)$
be the sets of mechanisms that are strategyproof, $r$-partially strategyproof, and LD-strategyproof in the setting $(N,M,q)$, respectively.
Then
\begin{eqnarray}
    \text{SP}(N,M,q)     & = & \bigcap_{r < 1} r\text{-PSP}(N,M,q), \label{EQ:SP_INTERSECT_PSP} \\
    \text{LD-SP}(N,M,q)    & = & \bigcup_{r > 0} r\text{-PSP}(N,M,q). \label{EQ:LDSP_UNION_PSP}
\label{EQ:PROP:SPECTRUM_LIMITS:LDSP}
\end{eqnarray}
\end{proposition}
\begin{proof}
Towards Equality (\ref{EQ:SP_INTERSECT_PSP}), observe that any utility function $u_i$ consistent with a strict preference order $P_i$ is also contained in \URBIr\ for some $r<1$.
A mechanism that is $r$-partially strategyproof for all $r<1$ therefore makes truthful reporting a dominant strategy for all utility functions.
This implies EU-strategyproofness.

Equality (\ref{EQ:LDSP_UNION_PSP}) follows
directly from Lemma~\ref{LEM:PSP_LDSP_EQUIV}. 
\end{proof}
In words, Proposition~\ref{PROP:SPECTRUM_LIMITS} shows that
there is no gap ``at the top'' between the least manipulable partially strategyproof mechanisms and those that are fully strategyproof, 
and there is no gap ``at the bottom'' between the most manipulable partially strategyproof mechanisms and mechanisms that are merely LD-strategyproof.
The degree of strategyproofness thus parametrizes the entire spectrum between full strategyproofness and the minimal LD-strategyproofness.

An important consequence of Proposition~\ref{PROP:SPECTRUM_LIMITS} is that, for a fixed setting, any LD-strategyproof mechanism is also $r$-partially strategyproof for some $r>0$.
On a technical level, this means that it suffices to verify LD-strategyproofness to show partial strategyproofness.
We use this in Section~\ref{SEC:APPLICATIONS} to prove partial strategyproofness of
the probabilistic serial mechanism. 
Furthermore, describing a mechanism as being LD-strategyproof ignores the additional information about incentive properties captured by partial strategyproofness, i.e., the parametric nature of the set of utility functions for which truthful reporting is guaranteed to be a dominant strategy under that mechanism.
\begin{remark}[Computability]
    \label{REM:COMPUTABILITY}
    For any mechanism and any setting, 
    the degree of strategyproofness is computable. 
    To show this, we provide a simple algorithm (Algorithm~\ref{ALG:FINDRHO} in Appendix~\ref{APP:DDSP_COMP}), 
    which exploits an equivalent formulation of $r$-partial strategyproofness in terms of a finite system of inequalities.%
    \footnote{Note that computability is not obvious because the definition of $r$-partial strategyproofness (Definition~\ref{DEF:PSP}) involves infinitely many utility functions.} 
    Computability distinguishes partial strategyproofness from other concepts
    for which computability has not been established, e.g.,
    the \emph{comparison of mechanisms by their vulnerability to manipulation}
    \citep{PathakSoenmez2013SchoolAdmissionsReformInChicagoAndEnglandComparingMechanismsByTheirVulnerabilityToManipulation},
    \emph{strategyproofness in the large} \citep{AzevedoBudish2019StrategyProofnessInTheLarge},
    and \emph{convex strategyproofness} \citep{Balbuzanov2016ConvexStrategyproofnessWithAnApplicationToTheProbabilisticSerialMechanism}.
    However, tractability is an issue; we can currently only compute the degree of strategyproofness for small settings because Algorithm~\ref{ALG:FINDRHO} considers all (exponentially many) preference profiles. 
    In Appendix~\ref{APP:DDSP_COMP}, we discuss approaches towards improving tractability.
\end{remark}
\subsection{Relationships with Other Notions of Strategyproofness}
\label{SEC:PSP:RELATION}
In this section, we explore the relationship between partial strategyproofness and other notions of strategyproofness. 
Proposition~\ref{PROP:SPECTRUM_LIMITS} already established full strategyproofness and LD-strategyproofness as the upper and lower limit concepts.
In the following, we describe the relationships with
weak SD-strategyproofness \citep{BogomolnaiaMoulin2001ANewSolutionToTheRandomAssignmentProblem},
convex strategyproofness \citep{Balbuzanov2016ConvexStrategyproofnessWithAnApplicationToTheProbabilisticSerialMechanism},
approximate strategyproofness \citep{Carroll2013AQuantitativeApproachToIncentivesApplicationToVotingRules},
and strategyproofness in the large \citep{AzevedoBudish2019StrategyProofnessInTheLarge}.

In their seminal work, \citet{BogomolnaiaMoulin2001ANewSolutionToTheRandomAssignmentProblem} used \emph{weak SD-strategyproofness} to describe
the incentive properties of the probabilistic serial mechanism.
\begin{definition}
\label{DEF:WSP}
A mechanism $\f$ is \emph{weakly SD-strategyproof} if,
for
all agents $i \in N$,
all preference profiles $(P_i,P_{-i}) \in \mathcal{P}^N$,
and all misreports $P_i' \in \mathcal{P}$,
the assignment vector $\f_i(P_i,P_{-i})$ stochastically dominates $\f_i(P_i',P_{-i})$ at $P_i$
whenever the two assignment vectors are comparable by stochastic dominance at $P_i$.
\end{definition}
Equivalently, this concept can be formulated in terms of expected utilities:
A mechanism is
weakly SD-strategyproof
if, for
all agents $i \in N$,
all preference profiles $(P_i,P_{-i}) \in \mathcal{P}^N$,
and all misreports $P_i' \in \mathcal{P}$,
there exists a utility function $u_i \in U_{P_i}$ such that
$\mathds{E}_{\f_i(P_i,P_{-i})}[u_i] \geq \mathds{E}_{\f_i(P_i',P_{-i})}[u_i]$.
Observe that $u_i$ can depend on $i$, $P_i$, $P_{-i}$, and $P_i'$,
which makes this requirement very weak. 
The slightly stronger incentive concept of \emph{convex strategyproofness} \citep{Balbuzanov2016ConvexStrategyproofnessWithAnApplicationToTheProbabilisticSerialMechanism} arises if $u_i$ may only depend on $i$ and $P_i$ but must be independent of $P_i'$ and $P_{-i}$.
\begin{definition}
\label{DEF:CSP}
A mechanism $\f$ is \emph{convex strategyproof} if,
for
all agents $i \in N$
and all preference orders $P_i\in \mathcal{P}$,
there exists a utility function $u_i \in U_{P_i}$ such that,
for all preferences $P_{-i} \in \mathcal{P}^{N\backslash\{i\}}$
and all misreports $P_i' \in \mathcal{P}$, we have
$\mathds{E}_{\f_i(P_i,P_{-i})}[u_i] \geq \mathds{E}_{\f_i(P_i',P_{-i})}[u_i]$.
\end{definition}
\citet{Balbuzanov2016ConvexStrategyproofnessWithAnApplicationToTheProbabilisticSerialMechanism} showed that the probabilistic serial mechanism is convex strategyproof and constructed a mechanism that is weakly SD-strategyproof but not convex strategyproof, which shows that the latter is a strictly stronger requirement.

While convex strategyproofness makes truthful reporting a dominant strategy for \emph{some} agents, 
the different notion of \emph{approximate strategyproofness} applies to \emph{all} agents but
only bounds their potential gain from misreporting by a small albeit positive amount. 
However, in ordinal domains, bounding these gains in a meaningful way is challenging because utilities are typically not comparable across agents. 
Nevertheless, one can formalize approximate strategyproofness for assignment mechanisms with the additional assumption that the agents' utility functions take values between 0 and 1  \citep{BirrellPass2011ApproximatelyStrategyproofVoting,Carroll2013AQuantitativeApproachToIncentivesApplicationToVotingRules,Lee2015EfficientPrivateAndEpsilonStrategyproofElicitationOfTournamentVotingRules}.
\begin{definition}
\label{DEF:ASP}
Given a setting $(N,M,q)$ and a bound $\varepsilon \in [0,1]$,
a mechanism $\f$ is
\linebreak
\emph{$\varepsilon$-approximately strategyproof}
if,
for all agents $i \in N$,
all preference profiles $(P_i,P_{-i}) \in \mathcal{P}^N$,
all misreports $P_i' \in \mathcal{P}$,
and all 
utility functions $u_i\in U_{P_i}$ with $u_i: M \rightarrow [0,1]$, we have 
$\mathds{E}_{\f_i(P_i,P_{-i})}[u_i] \geq \mathds{E}_{\f_i(P_i',P_{-i})}[u_i] - \varepsilon$. 
\end{definition}
Finally, \citet{AzevedoBudish2019StrategyProofnessInTheLarge} 
proposed \emph{strategyproofness in the large},
a concept which formalizes the intuition that agents
have a diminishing benefit from misreporting as markets get larger.
To formalize the sense in which markets \emph{get large}, we follow \citet{KojimaManea2010IncentivesInTheProbabilisticSerialMechanism} and consider a sequence of settings $(N^n,M^n,q^n)_{n\geq 1}$  with
a constant set of objects ($M^n=M$),
a growing agent population ($|N^n|=n$),
and growing capacities ($\min_{j\in M}q_j^n \rightarrow \infty$ as $n\rightarrow\infty$)
that satisfy overall demand ($\sum_{j\in M}q_j^n \geq n$).
\begin{definition}
\label{DEF:SPL}
For a finite set of utility functions $\{u^1,\ldots,u^K\}$ and a sequence of settings $(N^n,M^n,q^n)_{n\geq 1}$ with the above properties,
a mechanism is \emph{strategyproof in the large (SP-L)} if,
for all $\varepsilon > 0$,
there exists $n_{0} \in \mathds{N}$ such that, for all $n \geq n_0$,
no agent with a utility function from $\{u^1,\ldots,u^K\}$ can gain more than $\varepsilon$ by misreporting.%
\footnote{The original definition of strategyproofness in the large weakens strategyproofness further by also taking an \emph{interim perspective},
which assumes that agents are uncertain about the other agents' reports.
By omitting the interim perspective in our simplified Definition~\ref{DEF:SPL}, we obtain a more demanding version of SP-L.
In Theorem~\ref{THM:RELATIONS_TO_OTHER_CONCEPTS}, we show in what sense partial strategyproofness implies this more demanding version and therefore also the original definition of SP-L.}
\end{definition}
\begin{theorem}
\label{THM:RELATIONS_TO_OTHER_CONCEPTS}
The following hold:
\begin{enumerate}
\nosep
    \item
        \label{ITM:THM:RELATIONS_TO_OTHER_CONCEPTS:WSP_CSP}
        In any fixed setting $(N,M,q)$, $r$-partial strategyproofness for some $r>0$ implies convex strategyproofness and thus weak SD-strategyproofness.

        Conversely, 
        if $m=2$, then 
        full, 
        partial, 
        convex, 
        and weak SD-strategyproofness are equivalent;
        if $m\geq 3$, there exists a mechanism that is convex strategyproof (and thus weakly SD-strategyproof) but not $r$-partially strategyproof for any $r>0$.
    \item
        \label{ITM:THM:RELATIONS_TO_OTHER_CONCEPTS:ASP}
        In any fixed setting $(N,M,q)$, 
        $r$-partial strategyproofness for some $r>0$ implies $\varepsilon$-approximate strategyproofness for some $\varepsilon<1$;
        and for all $\varepsilon > 0$ there exists $r<1$ such that $r$-partial strategyproofness implies $\varepsilon$-approximate strategyproofness.

        Conversely,
        for all $\varepsilon > 0$, there exists an $\varepsilon$-approximately strategyproof mechanism that is not $r$-partially strategyproof for any $r>0$.
    \item
        \label{ITM:THM:RELATIONS_TO_OTHER_CONCEPTS:SPL}
        Fix a
        finite set of utility functions $\{u^1,\ldots,u^K\}$
        and a sequence $(N^n,M^n,q^n)_{n\geq 1}$ with
        $|N^n|=n$,
        $M^n=M$,
        $\sum_{j\in M}q_j^n \geq n$,
        and $\min_{j\in M}q_j^n \rightarrow \infty$ as $n\rightarrow\infty$.
        If the degree of strategyproofness of a mechanism $\f$ converges to $1$ as $n\rightarrow\infty$, then $\f$ is strategyproof in the large.

        Conversely, there exists a mechanism that is strategyproof in the large but not $r$-partially strategyproof for any $r>0$ in any of the settings $(N^n, M^n,q^n)_{n\geq 1}$.
\end{enumerate}
\end{theorem}
\begin{proof}
\emph{Statement~\ref{ITM:THM:RELATIONS_TO_OTHER_CONCEPTS:WSP_CSP}.}
Observe that $r$-partial strategyproofness implies convex strategyproofness because any utility function from \URBIr\ can take the role of $u_i$ in Definition~\ref{DEF:CSP},
and convex strategyproofness implies weak SD-strategyproofness.
Example~\ref{EX:CSP_NOT_PSP_MECHANISM} in Appendix~\ref{APP:PROOFS_RELATIONS} gives a mechanism that is convex strategyproof but violates upper invariance (and therefore partial strategyproofness).

\emph{Statement~\ref{ITM:THM:RELATIONS_TO_OTHER_CONCEPTS:ASP}.}
Let $\f$ be $r$-partially strategyproof for some $r>0$.
The proof of Lemma~\ref{LEM:PSP_LDSP_EQUIV} shows that any misreport under an $r$-partially strategyproof mechanism reduces the agent's assignment for the most preferred object for which the assignment changes by at least $\delta>0$. 
The mechanism is therefore $(1-\delta)$-approximately strategyproof. 

By Proposition~\ref{PROP:PSP_DD_EQUIVALENCE} in Appendix~\ref{APP:DDSP_COMP}, $r$-partial strategyproofness of $\f$ can be equivalently expressed in terms of the following finite system of inequalities:
for all $i \in N$, 
$(P_i,P_{-i}) \in \mathcal{P}^N$ (where $P_i:j_1 \succ\ldots\succ j_m$),
$P_i' \in \mathcal{P}$,
and $K \in \{1,\ldots,m\}$,
we have
\begin{equation}
    \sum_{k=1}^K     r^k \f_{i,j_k}(P_i,P_{-i})
    \geq \sum_{k=1}^K     r^k \f_{i,j_k}(P_i',P_{-i}).
\end{equation}
With $r>0$, we obtain
\begin{eqnarray}
    \sum_{k=1}^K     \f_{i,j_k}(P_i',P_{-i}) - \f_{i,j_k}(P_i,P_{-i}) & \leq &
        \sum_{k=1}^K     \left(1-r^k\right) \left(\f_{i,j_k}(P_i',P_{-i}) - \f_{i,j_k}(P_i,P_{-i})\right) \nonumber \\
    & \leq & \sum_{k=1}^K \left(1-r^k\right) \leq \sum_{k=1}^m \left(1-r^k\right).
    \label{EQ:UPPER_BOUND_GAIN_CONVERGENCE}
\end{eqnarray}
By choosing $r$ close to 1, the last term in (\ref{EQ:UPPER_BOUND_GAIN_CONVERGENCE}) can be made arbitrarily small. 
This implies a property that \citet{LiuPycia2016OrdinalEfficiencyFairnessAndIncentivesLargeMarkets} called \emph{$\varepsilon$-strategy-proofness}, which is equivalent to $\varepsilon$-approximate strategyproofness (by Theorem 1 in \citep{MennleSeuken2016_ECExtAbs_TheParetoFrontierForRandomMechanisms}). 
To see why the converse does not hold, refer to our proof of Statement~\ref{ITM:THM:RELATIONS_TO_OTHER_CONCEPTS:SPL}. 

\emph{Statement~\ref{ITM:THM:RELATIONS_TO_OTHER_CONCEPTS:SPL}.}
Observe that any utility function $u_i \in U_{P_i} $ satisfies \URBIr\ for some $r<1$.
Thus, we can choose  $\bar{r}<1$ large enough such that $u^k \in \URBI(\bar{r})$ for all $u^k \in \{u^1,\ldots,u^K\}$.
$\f$ is $\bar{r}$-partially strategyproof in all sufficiently large markets because the degree of strategyproofness of $f$ converges to 1 as markets get large (by assumption). Thus, in sufficiently large markets, truthful reporting is a dominant strategy for all agents with utility functions in  $\{u^1,\ldots,u^K\}$.

The converse does not hold:
Let $\f$ be a mechanism under which an agent gets 
some fixed object with probability $(n-1)/n$ and its reported last choice with probability $1/n$. 
This mechanism is $1/n$-approximately strategyproof in the $n^{\text{th}}$ setting and therefore strategyproof in the large,
but it is not even weakly SD-strategyproof in any setting.%
\footnote{While $\f$ is a valid counterexample,
it may appear artificial and inefficient.
To address this concern,
in Appendix~\ref{APP:PROOFS_RELATIONS},
we construct a mechanism that is 
ex-post efficient, 
anonymous, 
upper invariant, 
monotonic 
(i.e., ranking an object higher does not lead to a strictly lower assignment for that object), 
and $\varepsilon$-approximately strategyproof for a given $\varepsilon<1$, 
yet fails to be $r$-partially strategyproof for any $r>0$.
This highlights that partial strategyproofness and approximate strategyproofness are fundamentally different concepts: 
the former requires that there is no benefit from misreporting for a specific subset of the agents, while the latter requires that the benefit from misreporting is bounded for all agents.}
\end{proof}
Theorem~\ref{THM:RELATIONS_TO_OTHER_CONCEPTS} significantly expands the usefulness of the partial strategyproofness concept:
First, given a partially strategyproof mechanism, Statement~\ref{ITM:THM:RELATIONS_TO_OTHER_CONCEPTS:WSP_CSP} shows that weak SD-strategyproofness and convex strategyproofness are implied. 
Thus, for instance, partial strategyproof of the probabilistic serial mechanism (see Proposition~\ref{PROP:PS_PSP} in Section~\ref{SEC:APPLICATIONS:PS}) tightens prior results of \citet{BogomolnaiaMoulin2001ANewSolutionToTheRandomAssignmentProblem} and \citet{Balbuzanov2016ConvexStrategyproofnessWithAnApplicationToTheProbabilisticSerialMechanism}.

Second, by definition, $r$-partially strategyproof mechanisms make truthful reporting a dominant strategy for all agents whose utility functions satisfy \URBIr.
The definition of $r$-partial strategyproofness does not rule out that agents with utility functions violating \URBIr\ may benefit from misreporting.
However, Statement~\ref{ITM:THM:RELATIONS_TO_OTHER_CONCEPTS:ASP} shows that for those agents, their potential gain from misreporting is bounded and that this bound is tighter for greater $r$. 

The connection between partial and approximate strategyproofness allows a refinement of the straightforward and honest strategic advice to agents who participate in an $r$-partially strategyproof mechanism:
\emph{An agent is best off reporting truthfully if its preference intensities differ sufficiently
(i.e., the ratio between its normalized utilities for any two objects is at most $r$);
otherwise, the agent's potential gain from misreporting may be positive but it is quantifiably bounded (by $\varepsilon$ times the utility difference between its first and last choice).} 
Ideally, the values of $r$ and $\varepsilon$
can be determined by the mechanism designer for a given mechanism in a given setting.
We discuss computability of $r$ in Appendix~\ref{APP:DDSP_COMP};
regarding computability of $\varepsilon$ see \citep{MennleSeuken2016_ECExtAbs_TheParetoFrontierForRandomMechanisms}.
When the exact values of $r$ and $\varepsilon$ are not available,
upper or lower bounds may present a useful second best
(e.g., see \citep{Abaecherli2017AParametricProofSPLofPS_MScThesis} for an analytical lower bound on the degree of strategyproofness of the probabilistic serial mechanism).

Finally, \citet{AzevedoBudish2019StrategyProofnessInTheLarge}
developed strategyproofness in the large  to identify mechanisms for which incentives to misreport vanish in large markets. Convergence of a mechanism's degree of strategyproofness to 1 in large markets is an alternative way to formalize a similar  idea. It is reassuring to know that, by Statement~\ref{ITM:THM:RELATIONS_TO_OTHER_CONCEPTS:SPL},
 both concepts are consistent in the sense that, if a mechanism's degree of strategyproofness converges to 1 in large markets, then it is also strategyproof in the large. We use this result in our proof of Corollary~\ref{COR:PS_SPL}.%
\section{Applications}
\label{SEC:APPLICATIONS}
We now present two applications of our new partial strategyproofness concept.
In Section~\ref{SEC:APPLICATIONS:PS}, we prove partial strategyproofness of the probabilistic serial mechanism.
This yields the most demanding description of the incentive properties of this mechanism known to date, both in any finite setting
and in the limit as markets get large.
In Section~\ref{SEC:APPLICATIONS:BM}, we show how partial strategyproofness can be employed to distinguish two common variants of the Boston mechanisms by their incentive properties, a distinction that has remained elusive until now.
\subsection{New Insights about the Probabilistic Serial Mechanism}
\label{SEC:APPLICATIONS:PS}
The \emph{probabilistic serial (PS)} mechanism \citep{BogomolnaiaMoulin2001ANewSolutionToTheRandomAssignmentProblem} is one of the most well-studied mechanism for the random assignment problem. 
It uses the \emph{Simultaneous Eating} algorithm to determine an assignment:
All agents begin by consuming probability shares of their respective most-preferred objects at equal speeds.
Once an object's capacity has been completely consumed, the agents consuming this object move on to their respective next most-preferred objects and continue consuming shares of these.
This process continues until all agents have collected a total of 1 in probability shares, and these shares constitute their final assignments.

The next proposition shows that PS is partially strategyproof.
\begin{proposition}
\label{PROP:PS_PSP}
Given a setting $(N,M,q)$, PS is $r$-partially strategyproof for some $r>0$.
\end{proposition}
\begin{proof}
PS is LD-strategyproof by Theorem 3 in \citep{Cho2018ProabilisticAssignmentAnExtensionApproach}.
Therefore, in any fixed setting, it is $r$-partially strategyproof for some $r>0$ by Lemma~\ref{LEM:PSP_LDSP_EQUIV}.
\end{proof}
Regarding the incentive properties of PS, \cite{BogomolnaiaMoulin2001ANewSolutionToTheRandomAssignmentProblem} already showed that it is weakly SD-strategyproof (but not strategyproof), 
and \citet{Balbuzanov2016ConvexStrategyproofnessWithAnApplicationToTheProbabilisticSerialMechanism} strengthened their result by showing that it is convex strategyproof.
Since partial strategyproofness is strictly stronger than both properties, Proposition~\ref{PROP:PS_PSP} establishes the most demanding description of the incentive properties of PS in finite settings known to date. 
\begin{figure}%
\begin{center}%
\begin{tikzpicture}[scale=2]
    \draw[-] (-0.05,0) -- (3.0,0) coordinate (x axis);
    \draw[-] (0,-0.05) -- (0,1.0) coordinate (y axis);
    \draw (0.0,1.0) -- (-0.05,1.0);
    \draw (0.0, -0.15) node {0};
    \draw (-0.15,0) node {0};
    \draw (-0.15,1) node {1};
    \draw (0.5,-0.15) node {$3$};
    \draw (1,-0.15) node {$30$};
    \draw (1.5,-0.15) node {$60$};
    \draw (2.0,-0.15) node {$90$};
    \draw (2.5,-0.15) node {$120$};
    \draw (1.5,-0.35) node {$n$};
    \draw (-0.15,.5) node {$\rho$};
    \fill[fill=Dandelion] (.5-.1,0) -- (.5+.1,0) -- (.5+.1,.75) -- (.5-.1,.75);
    \fill[fill=Dandelion] (1-.1,0) -- (1+.1,0) -- (1+.1,.91) -- (1-.1,.91);
    \fill[fill=Dandelion] (1.5-.1,0) -- (1.5+.1,0) -- (1.5+.1,.95) -- (1.5-.1,.95);
    \fill[fill=Dandelion] (2-.1,0) -- (2+.1,0) -- (2+.1,.97) -- (2-.1,.97);
    \fill[fill=Dandelion] (2.5-.1,0) -- (2.5+.1,0) -- (2.5+.1,.98) -- (2.5-.1,.98);
    \draw (0.5,0.75+0.15) node {$0.75$};
    \draw (1.0,0.91+0.15) node {$0.91$};
    \draw (1.5,0.95+0.15) node {$0.95$};
    \draw (2.0,0.97+0.15) node {$0.97$};
    \draw (2.5,0.98+0.15) node {$0.98$};
\end{tikzpicture}
\hspace{1em}
\begin{tikzpicture}[scale=2]
    \draw[-] (-0.05,0) -- (3.0,0) coordinate (x axis);
    \draw[-] (0,-0.05) -- (0,1.0) coordinate (y axis);
    \draw (0.0,1.0) -- (-0.05,1.0);
    \draw (0.0, -0.15) node {0};
    \draw (-0.15,0) node {0};
    \draw (-0.15,1) node {1};
    \draw (0.6,-0.15) node {$4$};
    \draw (1.2,-0.15) node {$8$};
    \draw (1.8,-0.15) node {$12$};
    \draw (2.4,-0.15) node {$16$};
    \draw (1.5,-0.35) node {$n$};
    \draw (-0.15,.5) node {$\rho$};
    \fill[fill=Dandelion] (0.6-.1,0) -- (0.6+.1,0) -- (0.6+.1,.50) -- (0.6-.1,.50);
    \fill[fill=Dandelion] (1.2-.1,0) -- (1.2+.1,0) -- (1.2+.1,.67) -- (1.2-.1,.67);
    \fill[fill=Dandelion] (1.8-.1,0) -- (1.8+.1,0) -- (1.8+.1,.75) -- (1.8-.1,.75);
    \fill[fill=Dandelion] (2.4-.1,0) -- (2.4+.1,0) -- (2.4+.1,.80) -- (2.4-.1,.80);
    \draw (0.6,0.50+0.15) node {$0.50$};
    \draw (1.2,0.67+0.15) node {$0.67$};
    \draw (1.8,0.75+0.15) node {$0.75$};
    \draw (2.4,0.80+0.15) node {$0.80$};
\end{tikzpicture}%
\end{center}%
\caption{Plot of $\rho_{(N,M,q)}(\text{PS})$ for $m=3$ (left) and $m=4$ (right) objects, for varying numbers of agents $n$, and evenly distributed capacities $q_j = n/m$.}%
\label{FIG:RHO_PS}%
\end{figure}
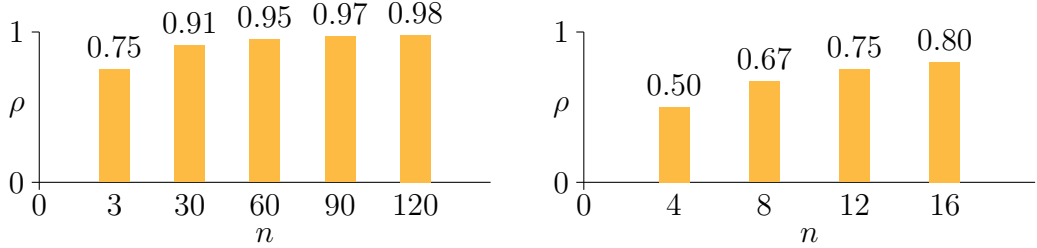

To see what degrees of strategyproofness PS  achieves, we have computed these values in various settings using Algorithm~\ref{ALG:FINDRHO} from Appendix~\ref{APP:DDSP_COMP}. Figure~\ref{FIG:RHO_PS} shows the results.
Observe that the values increase as the number of agents increases (while keeping the number of objects constant and evenly increasing capacities to satisfy demand). 
This suggests the hypothesis that the degree of strategyproofness of PS converges to 1 as the settings get large.%
\footnote{Note that, independent of our numerical results, \citet{KojimaManea2010IncentivesInTheProbabilisticSerialMechanism} presented a result about incentives under PS in large markets, which also suggests the same hypothesis but does not imply it.}
In subsequent work, \citet{Abaecherli2017AParametricProofSPLofPS_MScThesis} proved this hypothesis to be true.
We use his convergence result to obtain an elegant, parametric proof of the following corollary.
\begin{corollary}
\label{COR:PS_SPL}
    PS is strategyproof in the large.%
    \footnote{\citet{AzevedoBudish2019StrategyProofnessInTheLarge} first observed strategyproofness in the large for PS by invoking the result of \citet{KojimaManea2010IncentivesInTheProbabilisticSerialMechanism}.}
\end{corollary}
\begin{proof}
    Let $(N^n,M^n,q^n)_{n\geq 1}$ be a sequence of settings that get large as in Definition~\ref{DEF:SPL}.
    Theorem 1 of \citet{Abaecherli2017AParametricProofSPLofPS_MScThesis} implies that $\rho_{(N^n,M^n,q^n)}(PS)\rightarrow 1$ as $n \rightarrow\infty$.
    With this,
    Statement~\ref{ITM:THM:RELATIONS_TO_OTHER_CONCEPTS:SPL} of Theorem~\ref{THM:RELATIONS_TO_OTHER_CONCEPTS} implies that
     PS is strategyproofness in the large.
\end{proof}
The new insights about incentives under PS for both
finite settings and in large markets
highlight the usefulness of the partial strategyproofness concept.
In particular, they refine our understanding by providing the structure of the sets of utility functions (via the \URBIr\ domain restriction) 
for which truthful reporting is guaranteed to be a dominant strategy.
\begin{remark}
Our motivating example in the introduction considered PS in a setting with $3$ agents and $3$ objects with unit capacity.
Recall that agent 1 could benefit from misreporting if  $3/4 u_1(a) < u_1(b)$.
On the other hand, the degree of strategyproofness of PS in this setting is $3/4$ (see Figure~\ref{FIG:RHO_PS}),
which implies that agent 1 is best off reporting truthfully if  $3/4 u_1(a) \geq u_1(b)$.
Thus, in this setting, the set of utility functions for which PS makes truthful reporting a dominant strategy  is \emph{exactly} $\text{URBI}(3/4)$. 
Similarly, in a setting with 4 agents and 4 objects with unit capacity,
this set is  \emph{exactly}  $\text{URBI}(1/2)$.
However, in a setting with 5 agents and 5 objects with unit capacity, an analogous statement fails to hold
(see \citep{Abaecherli2017AParametricProofSPLofPS_MScThesis} for details on the settings with 4 and 5 agents). 
\end{remark}
\subsection{Comparing Variants of the Boston Mechanism}
\label{SEC:APPLICATIONS:BM}
Many school districts around the world employ school choice mechanisms to assign students to seats at public schools.
The \emph{Boston mechanism (BM)} \citep{AbdulkadirogluSoenmez2003SchoolChoiceAMechanismDesignApproach} is frequently used in the United States. 
Under BM, students apply to their respective first choices in the first round,
and schools accept applicants by priority.%
\footnote{School choice mechanisms take into account priorities \citep{AbdulkadirougluRothPathakSoenmez2006ChangingBostonSchoolChoiceStrategyproofnessAsEqualAccess},
which may depend on observable factors (e.g., grades, siblings, walk zones) and on lotteries (e.g., to break ties).}
If a school receives more applications than it has seats,
then it rejects applicants with the lowest priorities.
Rejected students enter the second round, where they apply to their respective second choices. 
Again, schools accept additional applicants according to priority and up to capacity,
and they reject the remaining applicants once all seats are filled.
This process continues with third, fourth, etc. choices until all seats are taken or all students have been assigned.

One notable aspect of BM is that a student who has been rejected by her first choice may find that all seats at her second choice have been taken in the first round as well.
Thus, when applying to her second choice in the second round, she effectively wastes one round where she could have competed for unfilled seats at other schools.
This makes ``skipping exhausted schools'' an obvious heuristic for manipulation.
A different variant, the \emph{adaptive Boston mechanism (ABM)}, is more common in Europe \citep{DurMennleSeuken2018FirstChoiceMaximalAndFirstChoiceStableSchoolChoiceMechanisms}.%
\footnote{See \citep{%
Alcalde1996ImplemetationOfStableSolutionsToMarriageProblems,%
Miralles2008TheCaseForTheBostonMechanism,%
Harless2019ImmediateAcceptanceWithOrWithoutSkipsComparingSchoolAssignmentProcedures,%
MennleSeuken2017TradeOffsInSchoolChoiceComparingDeferredAcceptanceTheNativeAndTheAdaptiveBostonMechanism,%
Dur2015TheModifiedBostonMechanism} 
for prior work that studied ABM.}
Under ABM, students apply to their \emph{best available choice} in every round.
Skipping exhausted schools is therefore no longer necessary.

Intuitively, this makes ABM less susceptible to manipulation than BM.
However, this difference is surprisingly challenging to formalize.
\citet{PathakSoenmez2013SchoolAdmissionsReformInChicagoAndEnglandComparingMechanismsByTheirVulnerabilityToManipulation} proposed a natural concept for comparing mechanisms by their \emph{vulnerability to manipulation}, but this concept fails to capture the difference between BM and ABM:
For the case of strict and fixed priorities, it indicates equivalence,
and for the case when priorities are determined by a single uniform lottery,
the comparison is inconclusive  \citep{DurMennleSeuken2018FirstChoiceMaximalAndFirstChoiceStableSchoolChoiceMechanisms}.
Interestingly, we can use partial strategyproofness to recover a meaningful distinction between BM and ABM.

To state this result formally, we require additional notation:
A \emph{priority order} $\pi_j$ is a strict order over agents, where $\pi_j: i \succ i'$ indicates that agent $i$ has priority over agent $i'$ at object $j$.
We denote by $\Pi$ the set of all possible priority orders.
A \emph{priority profile} $\pi=(\pi_j)_{j\in M} \in \Pi^M$ is a collection of priority orders of all objects, and $\pi$ is a \emph{single} priority profile if $\pi_j = \pi_{j'}$ for all $j,j'\in M$.
A \emph{priority distribution} $\mathds{P}$ is a probability distribution over priority profiles $\Pi^M$, and $\mathds{P}$ \emph{supports all single priority profiles} if $\mathds{P}[\pi] > 0$ for all single priority profiles $\pi$.
A \emph{school choice mechanism} $\varphi$ is a mapping $\varphi:\mathcal{P}^N \times \Pi^M \rightarrow X$ that selects a deterministic assignment based on a preference profile and a priority profile.
BM and ABM are examples of such school choice mechanisms.
Finally, for a priority distribution $\mathds{P}$,
we define $\varphi^\mathds{P}(P) = \sum_{\pi\in\Pi} \varphi(P,\pi) \mathds{P}[\pi]$.
Then $\f=\varphi^\mathds{P}$ is a random assignment mechanism,
and it captures the strategic situation of agents when priorities are drawn from $\mathds{P}$.

With this, we can formalize the distinction between mechanisms $\text{BM}^\mathds{P}$ and $\text{ABM}^\mathds{P}$.
\begin{proposition}
\label{PROP:BM_ABM_PSP}
Given a setting $(N,M,q)$
and a priority distribution $\mathds{P}$ that supports all single priority orders,
$\text{BM}^\mathds{P}$ and $\text{ABM}^\mathds{P}$ are upper invariant
and $\text{ABM}^\mathds{P}$ is swap monotonic, but 
$\text{BM}^{\mathds{P}}$ violates swap monotonicity if $m\geq 4$, $n\geq 4$, and $\sum_{j\in M}q_j=n$.
\end{proposition}
\begin{proof}
For all priority profiles $\pi\in\Pi^M$, upper invariance of $\text{BM}(\cdot,\pi)$ and $\text{ABM}(\cdot,\pi)$ is obvious.
Since $\text{BM}^{\mathds{P}}$ and $\text{ABM}^{\mathds{P}}$ are simply convex combinations of these mechanisms for different priority profiles, they inherit this property.
Next, observe that $\text{ABM}(\cdot,\pi)$ is \emph{monotonic} (i.e.,
for all $i\in N$,
$(P_i,P_{-i})\in\mathcal{P}^N$,
$P_i'\in N_{P_i}$ with $P_i:a\succ b$ but $P_i:b\succ a$, 
we have $\text{ABM}_{i,b}((P_i,P_{-i}),\pi) \leq \text{ABM}_{i,b}((P_i',P_{-i}),\pi)$),
and this property is again inherited by $\text{ABM}^{\mathds{P}}$. 
Lemma~\ref{LEM:ABM_SM_PROFILE_EXISTENCE} in Appendix~\ref{APP:PROOF_PSP_ABM_NOTPSP_BM} shows that $\text{ABM}_i^{\mathds{P}}(P_i,P_{-i}) \neq \text{ABM}_i^{\mathds{P}}(P_i',P_{-i})$ implies existence of a single priority profile $\pi$ with
$\text{ABM}_{i,b}((P_i,P_{-i}),\pi) < \text{ABM}_{i,b}((P_i',P_{-i}),\pi)$.
With this, monotonicity of $\text{ABM}^{\mathds{P}}$, and the fact that $\mathds{P}$ supports all single priority profiles, we obtain swap monotonicity of $\text{ABM}^{\mathds{P}}$.
Example~\ref{EX:BM_VIOLATES_SM} in Appendix~\ref{APP:PROOF_PSP_ABM_NOTPSP_BM} shows that $\text{BM}^{\mathds{P}}$ violates swap monotonicity in the respective settings.
\end{proof}
Proposition~\ref{PROP:BM_ABM_PSP} provides a formal justification for the intuition that ``ABM has better incentive properties than BM.''
If priorities are sufficiently random in the sense that all single priority profiles are possible, then $\text{ABM}^{\mathds{P}}$ is $r$-partially strategyproof for some $r>0$ while $\text{BM}^{\mathds{P}}$ has a degree of strategyproofness of 0. 
\begin{remark}
\label{REM:COARSE_PRIORITIES}
Proposition~\ref{PROP:BM_ABM_PSP} requires priorities to support all single priority profiles.
Various examples from practice meet this requirement, including middle school assignment in Bejing \citep{LaiSadoiuletJanvry2009TheAdverseEffectsOfParentsSchoolSelectionErrorsOnAcademicAchievementEvidenceFromTheBeijingOpenEnrollmentProgram},
inter-district assignment to elementary schools in Estonia  \citep{Lauri2014MIPEstoniaElementarySchools},
the supplementary round of high school assignment in New York City \citep{PathakSethuraman2011LotteriesInStudentAssignmentAnEquivalenceResult}, 
and secondary school assignment in Berlin \citep{BasteckKlausKuebler2018HowLotteriesInSchoolChoiceHelpToLevelThePlayingField}.

When priorities are coarse (e.g., based on siblings or walk-zones)
or even strict (e.g., based on GPA or entrance examinations), 
Proposition~\ref{PROP:BM_ABM_PSP} cannot be used directly to establish
partial strategyproofness of $\ABM^{\mathds{P}}$. 
To apply the partial strategyproofness concept in these cases, and more generally, to mechanisms with insufficient intrinsic randomness (and even deterministic ones),
one can take an \emph{interim perspective}
by considering the additional randomness that arises from the agents' uncertainty about the preference reports from the other agents
(see Section 2.8 of \citet{Mennle2016_thesis_TradeOffsBetweenStrategyproofnessAndEfficiencyOfOrdinalMechanisms} for a formal treatment of this approach).%
\footnote{This interim perspective is central to the definition of strategyproofness in the large \citep{AzevedoBudish2019StrategyProofnessInTheLarge}. 
In work subsequent to the present paper, \citet{DasguptaMishra2020OrdinalBayesianIncentiveCompatibilityInRandomAssignmentModel} used an interim perspective to study ordinal Bayesian incentive compatibility of random assignment mechanisms.}
\end{remark}
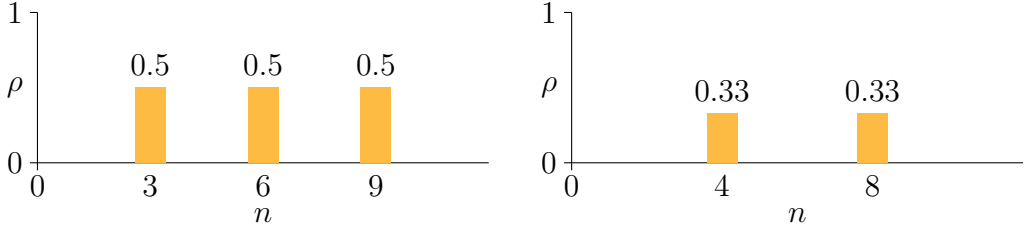
\begin{figure}%
\begin{center}%
\begin{tikzpicture}[scale=2]%
    \draw[-] (-0.05,0) -- (3.0,0) coordinate (x axis);
    \draw[-] (0,-0.05) -- (0,1.0) coordinate (y axis);
    \draw (0.0,1.0) -- (-0.05,1.0);
    \draw (0.0, -0.15) node {0};
    \draw (-0.15,0) node {0};
    \draw (-0.15,1) node {1};
    \draw (0.75,-0.15) node {$3$};
    \draw (1.50,-0.15) node {$6$};
    \draw (2.25,-0.15) node {$9$};
    \draw (1.5,-0.35) node {$n$};
    \draw (-0.15,.5) node {$\rho$};
    \fill[fill=Dandelion] (.75-.1,0) -- (.75+.1,0) -- (.75+.1,.5) -- (.75-.1,.5);
    \fill[fill=Dandelion] (1.5-.1,0) -- (1.5+.1,0) -- (1.5+.1,.5) -- (1.5-.1,.5);
    \fill[fill=Dandelion] (2.25-.1,0) -- (2.25+.1,0) -- (2.25+.1,.5) -- (2.25-.1,.5);
    \draw (0.75,0.50+0.15) node {$0.5$};
    \draw (1.5,0.50+0.15) node {$0.5$};
    \draw (2.25,0.50+0.15) node {$0.5$};
\end{tikzpicture}%
\hspace{1em}
\begin{tikzpicture}[scale=2]
    \draw[-] (-0.05,0) -- (3.0,0) coordinate (x axis);
    \draw[-] (0,-0.05) -- (0,1.0) coordinate (y axis);
    \draw (0.0,1.0) -- (-0.05,1.0);
    \draw (0.0, -0.15) node {0};
    \draw (-0.15,0) node {0};
    \draw (-0.15,1) node {1};
    \draw (1.0,-0.15) node {$4$};
    \draw (2.0,-0.15) node {$8$};
    \draw (1.5,-0.35) node {$n$};
    \draw (-0.15,.5) node {$\rho$};
    \fill[fill=Dandelion] (1.0-.1,0) -- (1.0+.1,0) -- (1.0+.1,.33) -- (1.0-.1,.33);
    \fill[fill=Dandelion] (2.0-.1,0) -- (2.0+.1,0) -- (2.0+.1,.33) -- (2.0-.1,.33);
    \draw (1.0,0.33+0.15) node {$0.33$};
    \draw (2.0,0.33+0.15) node {$0.33$};
\end{tikzpicture}
\end{center}%
\caption{Plot of $\rho_{(N,M,q)}(\text{ABM}^\mathds{U})$ for $m=3$ (left) and $m=4$ (right) objects, for varying numbers of agents $n$, and for $q_j = n/m$.}%
\label{FIG:RHO_ABM}%
\end{figure}
To obtain some insights regarding what degrees of strategyproofness ABM achieves, we focus on the  \emph{single uniform priority distribution} $\mathds{U}$ (i.e., the distribution that selects all single priority profiles with equal probability). Thus, for $\text{ABM}^\mathds{U}$,  
we have computed its degrees of strategyproofness in various settings.%
\footnote{The fastest known algorithm to compute the assignments under $\text{ABM}^\mathds{U}$ has exponential run-time (in contrast to PS, for which it is polynomial).
For this reason, we have only computed the degrees of strategyproofness of $\text{ABM}^\mathds{U}$  in settings with up to $9$ agents and up to $4$ objects (see Remark~\ref{REM:COMPUTABILITY} and Appendix~\ref{APP:DDSP_COMP} for a discussion of computability and tractability).}
The results are shown in Figure~\ref{FIG:RHO_ABM}. 
Observe that $\rho_{(N,M,q)}(\text{ABM}^\mathds{U})$ is significantly lower than $\rho_{(N,M,q)}(\text{PS})$ (Figure~\ref{FIG:RHO_PS}).
Furthermore, it appears to be constant as the number of agents increases (while keeping the number of objects constant and evenly increasing capacities to satisfy demand). 
At least for these small settings, our numbers support the intuition that ``PS has better incentive properties than $\ABM^\mathds{U}$.''%
\footnote{Note that we do not advocate the use of any mechanism solely on the basis of partial strategyproofness. 
Instead, such a decision must involve additional criteria like fairness and efficiency. Moreover, our results do not pertain to the controversy in the school choice literature that strategic reporting under the Boston mechanism can lead to ex-ante welfare gains \citep{Abdulkadirouglu2011}.}
\section{Conclusion}
\label{SEC:CONCLUSION}
In this paper, we have introduced partial strategyproofness, a new concept to understand the incentive properties of non-strategyproof assignment mechanisms.
This research is motivated by
the restrictive impossibility results pertaining to strategyproofness in the assignment domain
as well as the prevalence of non-strategyproof assignment mechanisms in practice.

In quasi-linear domains, such as auction problems, the monetary benefit from misreporting can serve as a quantifiable proxy for the extent to which agents care about strategizing.
However, such a proxy is more challenging to define in ordinal domains, where cardinal preferences are typically not comparable across agents.
The partial strategyproofness concept elegantly circumvents this problem:
It exploits the observation that whether an agent can manipulate an assignment mechanism is often driven by how close that agent is to being \emph{indifferent} between any two objects.
The \URBIr\ domain restriction separates agents by this criterion.
By requiring good incentives only on the restricted domain,
partial strategyproofness relaxes strategyproofness in a meaningful way and creates room for designing mechanisms that perform well on other dimensions. 

The partial strategyproofness concept strikes a unique balance between two conflicting goals:
It is strong enough to produce new insights yet weak enough to expand the mechanism design space. 
Regarding new insights, it allows us to provide honest and useful strategic advice to all agents: 
They are best off reporting their preferences truthfully if their
preference intensities 
differ sufficiently between any two objects; 
otherwise, if they are close to being indifferent between some objects,
then their potential gain from misreporting is quantifiably bounded in the sense of approximate strategyproofness. 
Regarding the expansion of the mechanism design space,
we have demonstrated that partial strategyproofness can be applied to
all mechanisms that satisfy the minimal requirement of lexicographic dominance strategyproofness.
In particular, these include some of the most important assignment mechanisms,
like probabilistic serial and variants of the Boston mechanism. 

An important open question is what values of $r$ are ``high enough'' to provide acceptable incentive guarantees.
We consider this an important and interesting subject of future research.
However, we do not believe that a universal answer to this question exists.
Rather, the appropriate degree of strategyproofness will depend on the particular mechanism design problem at hand.
For specific markets, it could be derived from survey data or revealed preference data (e.g., when preferences are expressed in terms of bidding points).
Such research could be complemented by laboratory experiments that aim to identify the role of the degree of strategyproofness in human decisions to manipulate non-strategyproof assignment mechanisms.

Despite this open question, partial strategyproofness can already guide design decisions.
For example, we can use it to compare non-strategyproof assignment mechanisms by their incentive properties (as we have illustrated by applying it to PS, ABM, and BM).
In addition, the parametric nature of partial strategyproofness enables new quantifiable trade-offs between incentives and other design desiderata.
For example, in \citep{MennleSeuken2017HybridMechanismsTradingOffStrategyproofnessAndEfficiencyOfRandomAssignmentMechanisms}, we have used it to identify the possible and necessary trade-offs between strategyproofness and efficiency that can be achieved via hybrid mechanisms.
Going forward, we are confident that the partial strategyproofness concept will be a useful addition to the mechanism designer's toolbox and that it will facilitate the study of non-strategyproof assignment mechanisms and the design of new ones.


\appendix
\section*{Appendix}
\section{Proof of Lemma~\ref{LEM:VARPHI_TILDE_PSP_BUT_MANIP} for Proposition~\ref{PROP:MAXIMALITY}}
\label{APP:VARPHI_TILDE_PSP_BUT_MANIP_LEMMA}
\begin{lemma}
\label{LEM:VARPHI_TILDE_PSP_BUT_MANIP}
The mechanism $\tilde\f$ constructed in the proof of Proposition~\ref{PROP:MAXIMALITY} is $r$-partially strategyproof and manipulable for agent $i$ with utility function $\tilde{u_i}$.%
\end{lemma}
\begin{proof}
Observe that, under $\tilde{\f}$, $i$'s assignment is independent of the other agents' preference reports, so the mechanism is strategyproof for all other agents except $i$.
Thus, we can guarantee $r$-partial strategyproofness of $\tilde{\f}$ by verifying that truthful reporting maximizes $i$'s expected utility for any preference order $P_i$ and any utility function $u_i\in U_{P_i}$ that satisfies \URBIr.

If $P_i : a \succ b$, then $i$'s assignment remains unchanged, unless $i$ claims to prefer $b$ to $a$ (i.e., $\hat{P}_i:b \succ a$).
If $a$ is $i$'s reported last choice under $\hat{P}_i$, then the change in $i$'s expected utility from the misreport is $- \delta_b u_i(a) + \delta_b u_i(b) \leq 0$.
If $\hat{d} \neq a$ is $i$'s reported last choice, then this change is
$    -\delta_a u_i(a) + \delta_b u_i(b) + (\delta_a - \delta_b) u_i(\hat{d})$.
Since $\delta_a < \delta_b$ by assumption, this value strictly increases if $i$ ranks its true last choice $d$ last.
Thus, $i$'s gain from misreporting is upper bounded by
\begin{eqnarray}
    & & -\delta_a u_i(a) + \delta_b u_i(b) + (\delta_a - \delta_b) u_i(d) \\
    & = & - \delta_a \left( u_i(a) - \min_{j \in M} u_i(j) \right) + \delta_b \left( u_i(b) - \min_{j \in M} u_i(j) \right).
\end{eqnarray}
This bound is guaranteed to be weakly negative if $\delta_a \geq r \delta_b$, where we use the fact that $u_i$ satisfies \URBIr.

Next, if $P_i : b \succ a$, then $i$'s assignment from truthful reporting stochastically dominates any assignment that $i$ can obtain by misreporting.

Finally, we need to verify that $i$ finds a beneficial misreport if its utility function is $\tilde{u}_i$.
Let $P_i'$ be the same preference order as $\tilde{P}_i$, except that $a$ and $b$ trade positions.
The change in $i$'s expected utility from reporting $P_i'$ is
\begin{eqnarray}
    & & -\delta_a \tilde{u}_i(a) + \delta_b \tilde{u}_i(b) + (\delta_a - \delta_b) \tilde{u}_i(d) \\
    & = & - \delta_a \left( \tilde{u}_i(a) - \min_{j \in M} \tilde{u}_i(j) \right) + \delta_b \left( \tilde{u}_i(b) - \min_{j \in M} \tilde{u}_i(j) \right),
\end{eqnarray}
where $d \neq b$ is $i$'s true last choice.
This change is strictly positive if $\delta_a < \tilde{r} \delta_b$.
\end{proof}
\section{Examples Used in the Proof of Theorem~\ref{THM:RELATIONS_TO_OTHER_CONCEPTS}}
\label{APP:PROOFS_RELATIONS}
\begin{example}[Convex strategyproofness does not imply upper invariance]
\label{EX:CSP_NOT_PSP_MECHANISM}
Consider a setting with one agent $i$ and three objects $a,b,c$ with unit capacity.
Suppose that ranking $b$ over $c$ leads to an assignment of $y_i=(0,1/2,1/2)$ for $a,b,c$, respectively,
and ranking $c$ over $b$ leads to $x_i=(1/4,0,3/4)$.
Without loss of generality, let $u_i$ be such that the utility for the last choice is $0$ and
the utility for the second choice is $1$.
Then $i$ can only benefit from misreporting in the following two cases:
\begin{enumerate}[1.]
\nosep
    \item $P_i:a\succ b \succ c$ and $u_i(a) > 2$,
    \item $P_i:c\succ b \succ a$ and $u_i(c) \in (1,2)$.
\end{enumerate}
The mechanism is therefore convex strategyproof but not upper invariant.
\end{example}
\begin{example}[$\varepsilon$-approximate strategyproofness does not imply $r$-partial strategyproofness]
\label{EX:ASP_NOT_PSP_MECHANISM}
For any $\beta \in [0,1]$,
let $h^{\beta}$ be the mechanism that
collects the agents' preference reports
and
then applies $\BM^{\mathds{U}}$ (the Boston mechanism with priorities determined by a single uniform lottery) with probability $\beta$
and applies $\text{RSD}$ (the random serial dictatorship mechanism with the order of dictators determined by a single uniform lottery) with probability $(1-\beta)$.
Ex-post efficiency of $h^\beta$ follows from Theorem 2 of \citep{MennleSeuken2017HybridMechanismsTradingOffStrategyproofnessAndEfficiencyOfRandomAssignmentMechanisms} and the fact that $\BM^{\mathds{U}}$ and $\text{RSD}$ are both ex-post efficient.
Similarly, $h^{\beta}$ inherits
upper invariance,
monotonicity,
and anonymity
from $\BM^{\mathds{U}}$ and $\text{RSD}$
(for a proof of  monotonicity of $\BM^{\mathds{U}}$, see Claim 9 in Section 3.B.2 of \citep{Mennle2016_thesis_TradeOffsBetweenStrategyproofnessAndEfficiencyOfOrdinalMechanisms}).
Next, observe that $\text{RSD}$ is $0$-approximately strategyproof and $\BM^{\mathds{U}}$ is trivially $1$-approximately strategyproof.
With this, $h^\beta$ is $\beta$-approximately strategyproof by Theorem 2 of \citep{MennleSeuken2016_ECExtAbs_TheParetoFrontierForRandomMechanisms}.
Finally, Example 2 of \citep{MennleSeuken2017HybridMechanismsTradingOffStrategyproofnessAndEfficiencyOfRandomAssignmentMechanisms} demonstrates that $h^{\beta}$ is not weakly SD-strategyproof in a setting with  $n=6$ agents and $m=6$ objects with unit capacity,
and $h^\beta$ is therefore not $r$-partially strategyproof for any $r>0$ in this setting.
\end{example}
\section{Example and Lemma Used in the Proof of Proposition~\ref{PROP:BM_ABM_PSP}}
\label{APP:PROOF_PSP_ABM_NOTPSP_BM}
\begin{example}[Violation of swap monotonicity by BM]
\label{EX:BM_VIOLATES_SM}
Consider a setting with
$n\geq 4$ agents $N=\{1,2,3,4,\ldots,n\}$,
$m\geq 4$ objects $M=\{a,b,c,d,\ldots\}$,
and capacities $q_j \geq 1$ such that
$\sum_{j\in M} q_j = n$.
Let agent $1$ have preference order $P_1 : a \succ b \succ c \succ d \succ \ldots$,
let agents $2$ through $q_a+q_c+q_d$ have preference order $P_\cdot : a \succ b \succ \ldots$,
and for all objects $j \in M\backslash\{a,b,d\}$, let there be $q_j$ agents with preference order
$P_\cdot : j \succ \ldots$.
Then
$\text{BM}^{\mathds{P}}_1(P_1,P_{-1}) = (x,y,0,1-x-y,0,\ldots)$
for some $x \in (0,1), y\in (0,1-x)$ because $\mathds{P}$ supports all single priority profiles.
However, if $i$ reports $P_i':a\succ c \succ b \succ d \succ \ldots$, then
$\text{BM}^{\mathds{P}}_1(P_1',P_{-1}) = (x,0,0,1-x,0\ldots)$.
Observe that agent 1's assignment changes but its assignment for $c$ remains unchanged, a contradiction to swap monotonicity.
\end{example}
\begin{lemma}
\label{LEM:ABM_SM_PROFILE_EXISTENCE}
Given a setting $(N,M,q)$,
for all priority distributions $\mathds{P}$,
all $i\in N$,
all $(P_i,P_{-i}) \in \mathcal{P}^N$,
all $P_i' \in N_{P_i}$ with $P_i:a\succ b$ but $P_i':b\succ a$,
if
\begin{equation}
    \text{ABM}^{\mathds{P}}_i(P_i,P_{-i}) \neq \text{ABM}^{\mathds{P}}_i(P_i',P_{-i}),
\end{equation}
then
there exists a priority profile $\pi^*$ such that
\begin{equation}
    \text{ABM}_{i,a}((P_i,P_{-i}),\pi^*) = \text{ABM}_{i,b}((P_i',P_{-i}),\pi^*) = 1.
\end{equation}
\end{lemma}
\begin{proof}
From
$\text{ABM}^{\mathds{P}}_i(P_i,P_{-i}) \neq \text{ABM}^{\mathds{P}}_i(P_i',P_{-i})$
it follows that there exists a priority profile $\pi^0$ with
$\text{ABM}_i((P_i,P_{-i}),\pi^0) \neq \text{ABM}_i((P_i',P_{-i}),\pi^0)$.
Let $\text{ABM}_{i,j}((P_i,P_{-i}),\pi^0)=1$ and $\text{ABM}_{i,j'}((P_i',P_{-i}),\pi^0)=1$, i.e., $i$ gets $j$ by reporting $P_i$ and $i$ gets $j'\neq j$ by reporting $P_i'$,
and let $K$ be the round in which $i$ applies to (or skips) $a$ when reporting $P_i$.

We differentiate 6 cases,
identify that the first 3 are impossible,
and construct single the priority profile $\pi^*$ for the other 3:
\begin{description}
\nosep
    \item[$\bm{P_i:j\succ a}$ or $\bm{P_i:j'\succ a}$:]
        Then upper invariance implies $j'=j$, a contradiction.
    \item[$\bm{j=b}$ or $\bm{j'=a}$:]
        Then monotonicity implies $j'=j$, a contradiction.
    \item[$\bm{P_i:b\succ j}$ and $\bm{P_i:b\succ j'}$:]
        The fact that $i$ gets neither $a$ nor $b$ when reporting $P_i$ or $P_i'$ means that both objects are exhausted by other agents at the end of round $K$.
        $i$ thus applies to at most one of them and skips the other.
        Since $i$ is rejected, the application process after round $K$ is the same under both reports.
        This implies $j=j'$, an contradiction.
    \item[$\bm{j=a}$ and $\bm{j'=b}$.]
        Then $i$ receives $a$ or $b$ in round $K$ under the different reports, respectively.
        Observe that the application process before this round is independent of whether $i$ reports $P_i$ or $P_i'$.
        For all $k\in \{1,\ldots,K-1\}$, let $N^k$ be the agents who receive their object in round $k$,
        and let $\pi^*_j$ be a priority order which,
        first, gives highest priority to all agents in $N_1$, then all agents in $N_2$, and so on,
        and second, ranks $i$ directly after the agents in $N_{K-1}$.
        Then the single priority profile $\pi^*$ where all objects have the same priority order $\pi^*_j$ is a single priority profile with $\text{ABM}_{i,a}((P_i,P_{-i}),\pi^*) = \text{ABM}_{i,b}((P_i',P_{-i}),\pi^*) = 1$.
    \item[$\bm{j = a}$ and $\bm{P_i: b \succ j'}$:]
        If $b$ is exhausted at the end of round $K-1$, then $i$ would skip $b$ when reporting $P_i'$ and still receive $a$.
        Thus, there is still capacity of $b$ available at the beginning of round $K$.
        Let $\pi^*$ be the same single priority profile as in the previous case.
        Then $\text{ABM}_{i,a}((P_i,P_{-i}),\pi^*) = \text{ABM}_{i,b}((P_i',P_{-i}),\pi^*) = 1$.
    \item[$\bm{j' = b}$ and $\bm{P_i: b \succ j}$:]
        This case is symmetric to the previous case but where the roles of $a$ and $b$ are inverted.
\end{description}
Thus, the single priority profile $\pi^*$ exists in all cases that do not imply contradictions.
\end{proof}
\section{Equivalence of Swap Monotonicity and Strategyproofness for Deterministic Mechanisms}
\label{APP:DET_MECH_SM_SP_EQUIV}
\begin{proposition}
\label{PROP:DET_MECH_SM_SP_EQUIV}
A deterministic mechanism $\f$ is strategyproof if and only if it is swap monotonic.
\end{proposition}
\begin{proof}
Since deterministic mechanisms are just special cases of random mechanisms, Theorem~\ref{THM:SP} applies:
A deterministic mechanism $\f$ is strategyproof if and only if it is swap monotonic, upper invariant, and lower invariant.
Thus, strategyproofness implies swap monotonicity (i.e., \emph{sufficiency} in Proposition~\ref{PROP:DET_MECH_SM_SP_EQUIV}).
For \emph{necessity}, observe that swap monotonicity implies upper and lower invariance for deterministic mechanisms:
If a swap (say from $P_i:a\succ b$ to $P_i':b\succ a$) affects an agent's assignment, then the assignment must change strictly for the two objects $a$ and $b$ that are swapped.
But under a deterministic mechanism, this change can only be from 0 to 1 or from 1 to 0.
The only possible changes are therefore the ones where an agent receives $a$ with certainty if it reports $P_i:a\succ b$ and receives $b$ with certainty if she reports $P_i':b \succ a$.
\end{proof}
\section{Comparing Mechanisms
by Vulnerability to Manipulation and
Degree of Strategyproofness}
\label{APP:CONSISTENCY_VULNERABILITY_DOSP}
The next proposition shows that the comparison of mechanisms by their vulnerability to manipulation and by their degrees of strategyproofness are \emph{consistent} but not equivalent.
\begin{proposition}
\label{PROP:DOSP_ISM_CONSISTENT}
For any setting $(N,M,q)$ and mechanisms $\f,\g$, the following hold:
\begin{enumerate}
    \setlength{\itemsep}{0pt}
    \item \label{ITM:VULNERABILITY_CONSISTENCY:VUL_IMPL_DOSP} If $\g$ is strongly as manipulable as $\f$, then $\rho_{(N,M,q)}(\f) \geq \rho_{(N,M,q)}(\g)$.
    \item \label{ITM:VULNERABILITY_CONSISTENCY:DOSP_IMPL_VUL} If $\rho_{(N,M,q)}(\f) > \rho_{(N,M,q)}(\g)$, and if $\f$ and $\g$ are comparable by the strongly as manipulable as relation, then $\g$ is strongly as manipulable as $\f$.
\end{enumerate}%
\end{proposition}
In Proposition~\ref{PROP:DOSP_ISM_CONSISTENT}, the strongly as manipulable as relation is extended to random assignment mechanisms as follows:
\begin{definition}
For a given setting $(N,M,q)$ and two mechanisms $\f, \g$, we say that $\g$ is \emph{strongly as manipulable as} $\f$ if,
for
all agents $i\in N$,
all preference profiles $(P_i,P_{-i}) \in \mathcal{P}^N$,
and all utility functions $u_i \in U_{P_i}$, the following holds:
If there exists a misreport $P_i'\in \mathcal{P}$ such that
\begin{equation}
    \mathds{E}_{\f_i(P_i,P_{-i})}[u_i] < \mathds{E}_{\f_i(P_i',P_{-i})}[u_i],
\end{equation}
then there exists a (possibly different) misreport $P_i'' \in \mathcal{P}$ such that
\begin{equation}
    \mathds{E}_{\g_i(P_i,P_{-i})}[u_i] < \mathds{E}_{\g_i(P_i'',P_{-i})}[u_i].
\end{equation}
\end{definition}
In words, $\g$ is strongly as manipulable as $\f$ if any agent who can manipulate $\f$ in a given situation can also manipulate $\g$ in the same situation.
\begin{proof}[Proof of Proposition~\ref{PROP:DOSP_ISM_CONSISTENT}]
\emph{Statement~\ref{ITM:VULNERABILITY_CONSISTENCY:VUL_IMPL_DOSP}.}
Observe that, if $\f$ is strongly as manipulable as $\g$, then any agent who can manipulate $\g$ also finds a manipulation to $\f$.
Thus, the set of utility functions on which $\g$ makes truthful reporting a dominant strategy cannot be larger than the set of utilities on which $\f$ does the same.
This in turn implies $\rho_{(N,M,q)}(\f) \geq \rho_{(N,M,q)}(\g)$.

\emph{Statement~\ref{ITM:VULNERABILITY_CONSISTENCY:DOSP_IMPL_VUL}.}
Observe that, if $\rho_{(N,M,q)}(\f) > \rho_{(N,M,q)}(\g)$, then there exists a utility function $\tilde{u}$ in $\URBI\left(\rho_{(N,M,q)}(\f)\right)$, which is not in $\URBI\left(\rho_{(N,M,q)}(\g)\right)$, and for which $\g$ is manipulable, but $\f$ is not.
Thus, $\f$ cannot be strongly as manipulable as $\g$, but the converse is possible.%
\end{proof}
\section{Discounted Dominance and Computability}
\label{APP:DDSP_COMP}
In this section, we define a new dominance notion we call \emph{$r$-discounted dominance}
and prove that the induced incentive concept,
\emph{$r$-discounted dominance strategyproofness},
is equivalent to $r$-partial strategyproofness.
We then show that this equivalence
allows us to devise a simple algorithm for computing the degree of strategyproofness. 
We also discuss computational challenges and potential remedies.
\subsection{Discounted Dominance}
\label{APP:DDSP_COMP:DD_SP}
The new dominance notion generalizes stochastic dominance but includes $r$ as a discount factor.
\begin{definition}
\label{DEF:DD}
For
a bound $r \in [0,1]$,
a preference order $P_i\in \mathcal{P}$ with $P_i: j_1 \succ \ldots \succ j_m$,
and assignment vectors $x_i, y_i$,
we say that \emph{$x_i$ $r$-discounted dominates $y_i$} \emph{at $P_i$}
if, for all ranks $K \in \{1,\ldots,m\}$, we have
\begin{equation}
    \sum_{k=1}^K r^k x_{i,j_k} \geq \sum_{k=1}^K r^k y_{i,j_k}.
    \label{EQ:D_DOMINANCE}
\end{equation}
\end{definition}
Observe that, for $r=1$, this is precisely the same as stochastic dominance.
However, for $r<1$, the difference in the agent's assignment for the $k^{\text{th}}$ choice is discounted by the factor $r^k$.
Analogous to stochastic dominance for SD-strategyproofness, we can use $r$-discounted dominance (\emph{$r$-DD}) to define the corresponding incentive concept.
\begin{definition}
\label{DEF:DD_SP}
Given a setting $(N,M,q)$ and a bound $r\in (0,1]$,
a mechanism $\f$ is
\emph{$r$-DD-strategyproof}
if,
for
all agents $i\in N$,
all preference profiles $(P_i,P_{-i}) \in \mathcal{P}^N$, and
all misreports $P_i' \in \mathcal{P}$,
$\f_i(P_i,P_{-i})$ $r$-discounted dominates $\f_i(P_i',P_{-i})$ at $P_i$.
\end{definition}
The next proposition yields the equivalence to $r$-partial strategyproofness.
\begin{proposition}
\label{PROP:PSP_DD_EQUIVALENCE}
Given a setting $(N,M,q)$ and a bound $r\in [0,1]$, a mechanism $\f$ is $r$-partially strategyproof if and only if it is $r$-DD-strategyproof.%
\end{proposition}
\begin{proof}
Given the setting $(N,M,q)$, we fix
an agent $i\in N$,
a preference profile $(P_i,P_{-i})\in \mathcal{P}^N$,
and a misreports $P_i'\in \mathcal{P}$.
The following claim establishes equivalence of the $r$-partial strategyproofness constraints and the $r$-DD-strategyproofness constraints for any such combination $(i,(P_i,P_{-i}),P_i')$ with $x = \f_i(P_i,P_{-i})$ and $y = \f_i(P_i',P_{-i})$.
\begin{claim}
\label{CLAIM:DD_EQUIVALENCE_TECHNICAL}
Given a setting $(N,M,q)$, a preference order $P_i \in \mathcal{P}$, assignment vectors $x,y$, and a bound $r \in [0,1]$, the following are equivalent:
\begin{enumerate}[A.]
    \setlength{\itemsep}{0pt}
    \item \label{ITM:DD_EQUIVALENCE_TECHNICAL:PSP} For all utility functions $u_i \in U_{P_i}\cap \URBIr$ we have $\sum_{j\in M} u_i(j) x_j \geq \sum_{j\in M} u_i(j) y_j$.
    \item \label{ITM:DD_EQUIVALENCE_TECHNICAL:DD} $x_i$ $r$-discounted dominates $y_i$ at $P_i$.
\end{enumerate}
\end{claim}%
\begin{proof}[Proof of Claim~\ref{CLAIM:DD_EQUIVALENCE_TECHNICAL}]
\emph{Sufficiency (\ref{ITM:DD_EQUIVALENCE_TECHNICAL:DD} $\Rightarrow$ \ref{ITM:DD_EQUIVALENCE_TECHNICAL:PSP}).}
Let $P_i : j_1 \succ \ldots \succ j_m$.
Assume towards contradiction that Statement~\ref{ITM:DD_EQUIVALENCE_TECHNICAL:DD} holds, but that for some utility function $u_i \in U_{P_i}\cap \URBIr$, we have
\begin{equation}
    \sum_{l=1}^m u_i(j_l) \left(x_{j_l} - y_{j_l}\right) < 0.
    \label{EQ:LEMMA_2:LESS_FOR_ALL}
\end{equation}
Without loss of generality, we can assume $\min_{j\in M} u_i(j) = 0$.
Let $\delta_{k} = x_{j_k} - y_{j_k}$ and let
\begin{equation}
    S(K) = \sum_{k=1}^K u_i(j_k) (x_{j_k} - y_{j_k}) = \sum_{k=1}^K u_i(j_k) \delta_k.
\end{equation}
By assumption, $S(m) = S(m-1) < 0$ (see (\ref{EQ:LEMMA_2:LESS_FOR_ALL})), so there exists a smallest value $K' \in \{1,\ldots,m-1\}$ such that $S(K') < 0$, but $S(k) \geq 0$ for all $k<K'$.
Using Horner's method, we rewrite the partial sum and get
\begin{eqnarray}
    S(K') & = & \sum_{k=1}^{K'} u_i(j_k) \delta_{k} = \left(\frac{S({K'}-1)}{u_i(j_{K'})} + \delta_{K'}\right) u_i(j_{K'}) \\
        & = & \left(\left(\frac{S({K'}-2)}{u_i(j_{{K'}-1})} + \delta_{{K'}-1}\right) \frac{u_i(j_{{K'}-1})}{u_i(j_{K'})} + \delta_{K'}\right) u_i(j_{K'}) \\
        & = & \label{EQ:HORNER_TERM_FOR_REPLACEMENT} \left( \left( \ldots \left( \delta_1 \frac{u_i(j_1)}{u_i(j_2)}+ \delta_2 \right) \frac{u_i(j_2)}{u_i(j_3)} + \ldots \right) \frac{u_i(j_{{K'}-1})}{u_i(j_{K'})} + \delta_{K'}\right) u_i(j_{K'}).
\end{eqnarray}
Since $u_i$ satisfies \URBIr, the fraction $\frac{u_i(j_{K'-1})}{u_i(j_{K'})}$ is bounded from below by $1/r$.
But since $S(K'-1) \geq 0$ and $u_i(j_{K'-1}) > 0$, we must have that
\begin{equation}
    \left(\frac{S(K'-2)}{u_i(j_{K'-1})} + \delta_{K'-1}\right) = \frac{S(K'-1)}{u_i(j_{K'-1})} \geq 0.
\end{equation}
Therefore, when replacing $\frac{u_i(j_{K-1})}{u_i(j_K)}$ by $1/r$ in (\ref{EQ:HORNER_TERM_FOR_REPLACEMENT}) we only make the term smaller.
By the same argument, we can successively replace all the terms $\frac{u_i(j_{k-1})}{u_i(j_k)}$ and obtain
\begin{equation}
    0 > S(K') \geq \left( \left( \ldots \left( \frac{\delta_1}{r} + \delta_2 \right) \frac{1}{r} + \ldots \right) \frac{1}{r} + \delta_{K'}\right) u_i(j_{K'}) = \frac{u_i(j_{K'})}{r^{K'}} \sum_{k=1}^{K'} r^k \delta_k.
\end{equation}
This contradicts $r$-discounted dominance of $x$ over $y$ at $P_i$, since
\begin{equation}
\sum_{k=1}^{K'} r^k \delta_k
    = \sum_{k=1}^{K'} r^k \left(x_{j_k}-y_{j_k}\right) \geq 0.
\end{equation}

\medskip
\emph{Necessity (\ref{ITM:DD_EQUIVALENCE_TECHNICAL:PSP} $\Rightarrow$ \ref{ITM:DD_EQUIVALENCE_TECHNICAL:DD}).}
Let $P_i : j_1 \succ \ldots \succ j_m$.
Assume towards contradiction that Statement~\ref{ITM:DD_EQUIVALENCE_TECHNICAL:PSP} holds, but $x$ does not $r$-discounted dominate $y$ at $P_i$, i.e., for some $K \in \{1,\ldots,m\}$, we have
\begin{equation}
    \sum_{k=1}^K r^k x_{j_k} < \sum_{k=1}^K r^k y_{j_k},
\label{EQ:DD_CONDITION_VIOLATION_STRICT}
\end{equation}
and let $K$ be the smallest rank for which Inequality (\ref{EQ:DD_CONDITION_VIOLATION_STRICT}) is strict.
Then the value
\begin{equation}
    \Delta = \sum_{k=1}^K r^k \left(y_{j_k} - x_{j_k}\right),
\end{equation}
is strictly positive.
Let $D\geq d > 0$, and let $u_i$ be the utility function defined by
\begin{equation}
    u_i(j_k) = \left\{
        \begin{array}{ll}
            D r^{k} , & \text{ if }k\leq K, \\
            d r^{k} , & \text{ if }K+1 \leq k \leq m-1, \\
            0, & k = m.
        \end{array}
    \right.
\end{equation}
This utility function satisfies \URBIr.
Furthermore, we have
\begin{eqnarray}
    \sum_{j \in M} u_i(j) x_{j} - \sum_{j \in M} u_i(j) y_{j}
    & = & \sum_{l=1}^m u(j_l) \left(x_{i,j_l} - y_{i,j_l}\right) \\
    & = & D \sum_{k=1}^K r^k \left(x_{j_k} - y_{j_k}\right) + d \sum_{k=K+1}^{m-1} r^k \left(x_{j_k} - y_{j_k}\right) \nonumber\\
    & \leq & - D \Delta + d.
\end{eqnarray}
For $d < D \Delta$, $\sum_{j \in M} u_i(j) x_{j} - \sum_{j \in M} u_i(j) y_{j}$ is strictly negative, a contradiction.
\end{proof}%
This concludes the proof of Proposition~\ref{PROP:PSP_DD_EQUIVALENCE}.
\end{proof}
Proposition~\ref{PROP:PSP_DD_EQUIVALENCE} generalizes the equivalence between EU-strategyproofness and SD-strategyproofness \citep{Erdil2014StrategyProofStochasticAssignment}.
Moreover, it yields an alternative definition of $r$-partial strategyproofness in terms of discounted dominance.
This shows that the partial strategyproofness concept integrates nicely into the landscape of existing incentive concepts, many of which are defined using dominance notions (e.g., SD-, weak SD-, LD-strategyproofness, and sure thing dominance strategyproofness \citep{AzizBrandBrill2013OnTheTradeoffBetweenEconomicEfficiencyAndStrategyProofnessInRandomizedSocialChoice}).
\subsection{Computability}
\label{APP:DDSP_COMP:COMP}
The dominance interpretation from Proposition~\ref{PROP:PSP_DD_EQUIVALENCE} enables a computational approach to partial strategyproofness:
Recall that Definition~\ref{DEF:PSP} of $r$-partial strategyproofness imposes inequalities for all utility functions within the set \URBIr.
This set is infinite, which makes an algorithmic verification of $r$-partial strategyproofness infeasible via its original definition.
However, by the equivalence from Proposition~\ref{PROP:PSP_DD_EQUIVALENCE}, it suffices to verify that all (finitely many) constraints for $r$-discounted dominance strategyproofness are satisfied (i.e., the Inequalities (\ref{EQ:D_DOMINANCE}) from Definition~\ref{DEF:DD}).
This yields Algorithm~\ref{ALG:FINDRHO} to compute the degree of strategyproofness of any mechanism in any setting 
(provided that the mechanism $f$ itself is computable).

\begin{algorithm}
\SetAlgoLined
\KwIn{Mechanism $\f$, setting $(N,M,q)$}
\KwOut{Degree of strategyproofness $\rho_{(N,M,q)}(\f)$}

$r\leftarrow 1$

\For{all preference profiles $P \in \mathcal{P}^N$}{
    \For{all agents $i \in N$}{

        let $j_1,\ldots,j_m\in M$ be objects such that $P_i:j_1\succ\ldots\succ j_m$

        \For{all misreports $P_{i}'\in \mathcal{P}, P_{i}'\neq P_i$}{
            \For{all ranks $K\in\{1, \ldots,m\}$}{
                define polynomial $p(\cdot):s\mapsto \sum_{k=1}^K s^k\left( f_{i,j_k}\left(P_i,P_{-i}\right) - f_{i,j_k}\left(P_i',P_{-i}\right)\right)$
                
                $r' \leftarrow \max \left\{\overline{s}\in[0,1]:
                    p(s)\geq 0
                    \text{ for all }s\in [0,\overline{s}]
                \right\}$
                
                $r\leftarrow \min\left(
                    r,                
                    r'     
                \right)$
            }
        }
    }
}
\Return $r$
\caption{Find degree of strategyproofness}
\label{ALG:FINDRHO}
\end{algorithm}
The algorithm begins by optimistically assuming $r=1$. 
It then iterates over all possible combinations of 
preference profiles, 
agents, 
misreports, 
and ranks. 
For each combination, it checks whether the current $r$ must be corrected downwards to avoid a violation of the respective constraint for discounted dominance from Definition \ref{DEF:DD}. 
Specifically, 
we define the polynomial $p(\cdot)$ to represent the respective constraint. 
Next, we determine the largest bound $r'$ such that $p(s)\geq 0$ for all $s\in[0,r']$, 
and we reduce $r$ if $r'$ is smaller than $r$.%
\footnote{Since $p(0) = 0$ by construction, $r'$ is well-defined. 
Determining $r'$ requires an analysis of the roots of $p(\cdot)$. 
Finding roots of polynomials with rational coefficients is possible in polynomial time via the LLL algorithm \citep{LLL1982FactoringPolynomials}.}  
This ensures that all discounted dominance constraints checked so far are satisfied for the current $r$ but not for any larger values. 
Therefore, the value of $r$ when the algorithm terminates is the degree of strategyproofness of the mechanism $\f$ in the setting $(N,M,q)$. 

We have used Algorithm~\ref{ALG:FINDRHO} to compute the degrees of strategyproofness of PS and $\text{ABM}^{\mathds{U}}$ in a number of small settings in Section~\ref{SEC:APPLICATIONS}.
The limited setting sizes are owed to the exponential number of preference profiles and misreports over which the algorithm iterates.
This issue is further exacerbated by the fact that computing the assignment matrix for even a single preference profile can be a computationally expensive problem for some mechanisms
(e.g., \citet{Aziz2013ComplexityRSD} proved that computing the assignment resulting from random serial dictatorship is $\#P$-hard, and for $\text{ABM}^{\mathds{U}}$ the fastest known algorithm has exponential runtime). 
For these reasons, Algorithm~\ref{ALG:FINDRHO} is only practical in small settings and tractability may be an issue in larger settings.

To address the problem of tractability, we observe that most applications come with additional restrictions,
such as fairness requirements,
admissible types of mechanisms,
or limitations on preference reports.
Exploiting these restrictions provides avenues for future research.
For example, axioms like neutrality, symmetry, anonymity, or non-bossiness impose equality constraints.
These constraints can be exploited to reduce the number of preference profiles that must be considered.
Suppose that we restrict attention to anonymous mechanisms;
then,
for every preference profile we consider,
we can skip all other preference profiles that are equivalent to the first profile up to a permutation of the agents.

A second approach towards improving tractability arises from local sufficiency.
Even though full strategyproofness requires that none of the $\left(m!-1\right)$ misreports improve the agent's assignment,
it actually suffices to verify this for the $\left(m-1\right)$ misreports from the neighborhood of the agent's true preference order \citep{Carroll2012WhenAreLocalIncentiveConstraintsSufficient}.
The same is true for LD-strategyproofness, the lower limit concept of partial strategyproofness \citep{Cho2016IncentivePropertiesForOrdinalMechanisms}.
In \citet{MennleSeuken2017LocalSufficiencyForPartialStrategyproofness}, 
we have shown that $r$-\emph{local} partial strategyproofness%
\footnote{%
$r$-local partial strategyproofness requires that,   
for each agent whose utility function satisfies \URBIr, 
truthful reporting dominates all misreports from the \emph{neighborhood} of its true preference order.}
implies $r^2$-partial strategyproofness and that this bound is tight (i.e., $r$-local partial strategyproofness does not imply $r^{2-\varepsilon}$-partial strategyproofness for any $\varepsilon>0$).
Verifying that a mechanism is $r$-locally partially strategyproof thus yields a lower bound $r^2$ on its degree of strategyproofness but only requires checking $(m-1)$ misreports instead of $(m!-1)$.
\end{document}